\documentclass[12pt]{iopart}
\usepackage{graphicx}
\usepackage{natbib}

\usepackage{aas_macros}
\usepackage{txfonts}

 \usepackage{color}                                  %
 \definecolor{gray}{rgb}{0.6,0.6,0.6}                %
 \definecolor{green}{rgb}{0,0.6,0}                   

\begin{document}

\title{Circular geodesics and thick tori around rotating boson stars}

  \author{Z. Meliani$^{1}$,
          F. H. Vincent$^{2,3}$, 
          P. Grandcl\'ement$^{1}$, 
          E. Gourgoulhon$^{1}$, 
          R. Monceau-Baroux$^{1}$,  
          O. Straub$^{1}$}
\address{$^{1}$ LUTH, CNRS UMR 8102, Observatoire de Paris, Universit\'e Paris Diderot, 92190 Meudon, France\\
        $^{2}$ Nicolaus Copernicus Astronomical Center, ul. Bartycka 18, 00-716, Warszawa, Poland \\
        $^{3}$ LESIA, CNRS UMR 8109, Observatoire de Paris, Universit\'e Pierre et Marie Curie, Universit\'e Paris Diderot, 92190 Meudon, France}
\ead{zakaria.meliani@obspm.fr}
\vspace{10pt}
\begin{indented}
\item[]July 2015
\end{indented}

%

\begin{abstract}

  {Accretion disks play an important role in the evolution of their relativistic inner compact objects. The emergence of a new generation of interferometers will allow to resolve these accretion disks and provide more information about the properties of the  central gravitating object. Due to this instrumental leap forward it is crucial  to investigate the accretion disk physics near various types of inner compact objects now to deduce later constraints on the central objects from observations. A possible candidate for the inner object is the boson star.  Here, we will try to analyze the differences between accretion structures surrounding boson stars and black holes.}
  {We aim at analysing the physics of circular geodesics around boson stars and study simple thick accretion tori (so-called Polish doughnuts) in the vicinity of these stars.}
  {We realize a detailed study of the properties of circular geodesics around boson stars. We then perform a parameter study of thick tori with constant angular momentum surrounding boson stars. This is done using the boson star models computed by a code constructed with the spectral solver library KADATH.  }
  {We demonstrate that all the circular stable orbits are bound. In the case of a constant angular momentum torus, a cusp in the torus surface exists only for boson stars with a strong gravitational scalar field. Moreover, for each inner radius of the disk, the allowed specific angular momentum values lie within a constrained range which depends on the boson star considered.}
  {We show that the accretion tori around boson stars have different characteristics than in the vicinity of a black hole. With future instruments it could be possible to use these differences to constrain the nature of compact objects.}
\end{abstract}

%

\section{Introduction}
Accretion tori may be present in many astrophysical objects, e.g. microquasars and galactic nuclei. On the one hand they play a fundamental role in the evolution of their environment by converting gravitational potential energy into heat and radiation. On the other hand tori are subjected to  the gravitational field of the central compact object. Therefore, studying accretion structures like tori is essential to understand the physics and nature of the inner object.

Many authors have been investigating the physics of  accretion tori around black holes in the framework of general relativity with various theoretical models \citep[e.g.][]{Abramowiczetal78, Komissarov06, Font&Daigne02} and with extensive numerical simulations that range from two  to three dimensions  \citep[e.g.][]{Dibiet12, Fragileetal14, McKinneyetal14}. Analytic torus  models have recently been used to calculate  images of the accretion flow including  the black hole silhouettes \citep{Straubetal12, Vincentetal15}. These may be used to deduce the mass and spin of the central object.

Such studies are mainly motivated by the advent of near-future interferometric instruments which will open a new window on the observational study of strong-field gravity by giving reach to $10~\mu$as-scale physics, which corresponds to the angular size of the supermassive compact object Sgr~A* at the centre of the Milky Way or to the angular size of the compact object M87* at the centre of the galaxy M87. The GRAVITY second-generation Very Large Telescope Interferometer beam combiner \citep{eisenhauer11} will see its first light at the end of 2015 and will reach an astrometric precision of $10~\mu$as in the near infrared at the Galactic centre. Before 2020, the Event Horizon Telescope \citep[EHT,][]{doeleman09} will reach $20~\mu$as resolution in the sub-millimetre both at Sgr~A* and M87*, allowing to get images of the vicinity of these compact objects at Kerr event-horizon scale.

Until now, most studies were carried out using a black hole as the source of gravity. An alternative to the black hole much discussed in the literature is the \emph{boson star}, which is a self-gravitating stationary configuration of some massive complex scalar field. The concept of boson stars goes back to \citet{Bonazzola&Pacini66} and \citet{Kaup68}. These ``stars'' are characterized by a strong gravitational field and a vanishing (or very low) electromagnetic emissivity, which makes them act as ``black hole mimickers''. More about boson stars can be found in the review of \citet{Leibling&Palenzuela12}. Near a boson star, gravitation has specific features and thus orbital properties of gas, stars and radiation are different from scenarios where a black hole is the central compact object. It is crucial to study the differences between the two cases. Considering a boson star rather than a black hole can change the circular geodesics, the shape of the accretion torus and the radiation propagation. We need to investigate the imprint of boson stars on the nearby medium and the possible observable differences as compared to black holes. This is why it is important to study accretion thin and thick disks around boson stars. As in the black hole case,
we could build the multi-wavelength silhouette from the surrounding accretion disk and compare
it to near-future EHT data of Sgr~A* or M87*. Contrary to black holes, boson stars have a non-singular spacetime metric and do not possess any event horizon. Electromagnetic radiation can therefore arise from very close to the boson star centre. This fundamental difference may lead to particular observable features. The main goal of this paper and our near-future work is to
study these possible observable differences in detail. Indeed, with the current observations, it is possible to resolve the innermost accretion region of Sgr~A* \citep{Doelemanetal08}. However, we note that this size does not correspond to the size of the compact object itself. Moreover, the properties of the inner object such as its mass and predicted cosmological evolution are model-dependent \citep{Kormendyetal13}.

The viability of a supermassive boson star as an alternative to supermassive
black holes at the center of galaxies and in particular for Sgr A* has been 
investigated in great detail by Torres et al. (2000). They show that
the dynamical data provided at the time of writing were nicely fitted 
by a boson star, in much the same way as by a Kerr black hole. The problem
is actually to be able to differentiate between the two objects that can both
account for current observable constraints. 
The development of new high-resolution instruments like the EHT might allow us to check if boson stars exist
in the next decade.

Boson star masses can reach values comparable to supermassive black holes \citep{torres00}. The mass of the star is thereby inversely proportional to the mass of the constituting boson  and it depends on the potential that describes the boson self-interaction \citep{Grandclementetal14}. In the simplest case of a vanishing self-interaction (free scalar field) considered in this paper, the gravitational field generated by a stationary rotating boson star is characterized by two constants: an angular frequency $\omega$, which is a positive quantity that must be lower than the Compton angular velocity of the boson, and the rotational quantum number $k$, which is a non-negative integer.

In order to study complex phenomena such as the evolution of accretion tori in the vicinity of boson stars, we need first to understand the properties of circular geodesics around rotating and non-rotating boson stars and the characteristics of accretion tori.  In this paper, we will revisit the analysis of the innermost stable circular orbit and bound orbit in the context of boson stars.  We will analyze the orbit physics and apply these studies to the case of a thick torus with constant angular momentum. We should notice, that the study of the geodesics around non-rotating spherically symmetric boson stars has been performed by \citep{Diemeretal13,Macedoetal13}.

\section{Boson stars}
Throughout this paper, we consider rotating boson star models that have been computed by \citet{Grandclementetal14} by solving the Einstein-Klein-Gordon equations using a numerical code constructed upon the KADATH library \citep{Granclementetal10}. These objects consist of stationary and axisymmetric spacetimes generated by a massive complex scalar field $\Phi$ minimally coupled to gravity. Boson stars are obtained by choosing the following ansatz for $\Phi$, in terms of spherical coordinates $(t,r,\theta,\varphi)$:
\begin{equation}
\label{e:Phi_ansatz}
\Phi = \phi\left(r, \theta\right) \exp\left[i\left(\omega t-k\varphi\right)\right],
\end{equation}
where $k$ is a non-negative integer, called the \emph{rotational quantum number} and $\omega$ is a positive real constant, called hereafter the \emph{boson field pulsation}. The ansatz (\ref{e:Phi_ansatz}) leads to stationary and axisymmetric spacetimes. The integer $k$ sets the boson star total angular momentum $J$ via the following quantization law \citep{SchunckM96}:
\begin{equation}
\label{e:J_k}
J = k \hbar \mathcal{N},
\end{equation}
where $\mathcal{N}$ is the total number of bosons constituting the star. For $k=0$, the boson star is not rotating ($J=0$), the field $\phi$ depends only on $r$, and the spacetime metric is static and spherically symmetric. For $k \geq 1$, the boson star is rotating ($J>0$), the field $\phi$ is axisymmetric and has the topology of a torus, and the spacetime is stationary and axisymmetric. In this paper we consider only boson stars for which the scalar field is not
self-interacting (free field); they are named {\it mini boson stars} in the literature.

Under the above assumptions, a given boson star depends only on the choice of $k$ and $\omega$. One can show that $\omega$ must be smaller than the boson Compton angular frequency $\omega_{\rm b} := m c^2 / \hbar$ \citep[see][for details]{Grandclementetal14} and that the lower $\omega$ is, the more relativistic is the configuration. Besides, as shown by Eq.~(\ref{e:J_k}), the angular momentum becomes larger as $k$ increases. Table~\ref{Tab_KOMEGA} lists the various models of boson stars considered in this paper with the associated global quantities: the ADM (Arnowitt, Deser, Misner) mass $M$ and the reduced angular momentum $a/M := J/M^2$. Those configurations have been chosen to be as illustrative as possible. In particular, for each $k$, we consider the most relativistic configuration obtained in \citet{Grandclementetal14}. Let us recall that the configurations with $k=0$ are non-rotating (hence $a=0$) and spherically symmetric. We also note that most of the rotating boson stars have $a/M >1$. Actually, as soon as $k\not=0$, boson stars are fast rotators, as can be seen from Eq.~(\ref{e:J_k}). This is in sharp contrast with black holes, for which $a/M$ can vary smoothly from $0$ to $1$, the latter being the extreme Kerr limit.

\begin{table}
\caption{Models of boson stars considered in this article. Each star is labelled by its two characteristic numbers $k$ and $\omega$, with $\omega$ given in units of the Compton angular frequency of the boson, $\omega_{\rm b} = m/\hbar$, $m$ being the boson mass. The displayed quantities are the ADM mass $M$, expressed in in units of $m_{\rm P}^2 / m$,  where $m_{\rm P}$ is the Planck mass (see Sec.~\ref{units}), the reduced angular momentum $a := J/M$ and the radius of the innermost circular orbit $r_{\rm ICO}$.
}
$$
\begin{array}{lllll}
  \hline
   \noalign{\smallskip}
   k & \omega \ [m/\hbar]& M \ [m_{\rm P}^2/m] & a/M & r_{\rm ICO}/M\\
   \hline
   \noalign{\smallskip}
   0 & 0.7676 & 0.4898 & 0. & 0.0\\
   0 & 0.8117 & 0.6119 & 0. & 0.0\\
   0 & 0.9762 & 0.3611 & 0. & 0.0\\
   1 & 0.66 & 1.1621 & 0.8590 &0.55\\
   1 & 0.8 & 1.30779 & 0.8021 &1.208\\
   1 & 0.9 & 1.1186 & 0.9210 &2.78\\
   2 & 0.535 & 2.14889 & 0.9900 & 0.28\\
   2 & 0.8 & 2.01642 & 1.0509 &1.95\\
   2 & 0.9 & 1.60747 & 1.2838 &4.43\\
   3 & 0.505 & 3.46261 & 1.0057 &0.375\\
   3 & 0.8 & 2.69558 & 1.1855 &2.4\\
   3 & 0.9 & 2.0716 & 1.4958 &5.35\\
   4 & 0.56 & 4.49286 & 1.0402&0.153\\
   4 & 0.8 & 3.35477 & 1.2736 &2.7\\
   4 & 0.9 & 2.51739 & 1.6423 &6.057\\
   \hline
\end{array}
$$
\label{Tab_KOMEGA}
\end{table}

As in \citet{Grandclementetal14}, we adopt the quasi-isotropic coordinates $\left(t,r,\theta,\varphi\right)$, with line element defined as,
\begin{equation}\label{eq1_1}
{\rm d}s^2=-\alpha^2{\rm d}t^2+A^2\left({\rm d}r^2+r^2{\rm d}\theta^2\right)+B^2\,r^2\,\sin^2\theta\,\left({\rm d}\varphi+\beta^{\varphi}\,{\rm d}t\right)^2\,,
\end{equation}
where $\alpha$ is the lapse function and $\beta=\left(0,0,\beta^{\varphi}\right)$ is the shift vector. The quantities $A$, $B$, $\alpha$ and $\beta^\varphi$ depend only on $\left(r,\theta\right)$.

In this paper, once the geometry is known, we construct stationary axisymmetric thick tori with constant angular momentum around boson stars. We neglect the self-gravity and the fluid is defined with a barotropic equation of state $P(\epsilon)$ and with the stress-energy tensor
\begin{equation}\label{eq1_2}
T_{\mu\,\nu}=\rho\,h\,u_{\mu}\,u_{\nu}+P\,g_{\mu\,\nu}\,,
\end{equation}
where $\rho$ is the rest-mass density, $h=\left(1+\epsilon + \frac{P}{\rho}\right)$ is the specific enthalpy, $P$ is the pressure, $\epsilon$ is the specific internal energy, and $u^{\mu}$ is the fluid 4-velocity.

\section{Units}\label{units}
We use units in which the speed of light $c=1$ and the gravitation constant $G=1$. In this section only, for the sake of clarity, $c$ and $G$ factors will appear.

Following \citet{Grandclementetal14}, a natural mass scale that appears in the context of boson stars is
\begin{equation}
\mathcal{M}=\frac{m_{\rm P}^2}{m} = \frac{\hbar \,c}{G \, m}
\end{equation}
where $m_{\rm P}^2=\hbar c/G$ is the Planck mass squared and $m$ is the boson mass. The boson star total gravitational mass (ADM mass) is denoted $M$\footnote{Be careful not to confuse the elementary boson mass $m$ and the total boson star mass $M$.} and is expressed in units of $\mathcal{M}$ in Table~\ref{Tab_KOMEGA}. Let us note that $M$ can be seen as a scaling factor in much the same way as the total mass of a black hole in Kerr spacetime. However here, the bigger $m$ is, the smaller the unit length becomes. For one given value of $m$, the boson star mass $M$ can vary by a factor of a few when varying $\omega$ and $k$ \citep{Grandclementetal14}. Any specified compact object mass can be obtained with a mini boson star, provided a suitable boson exists, which we assume in the following. We note that for a given value of $m$, a broader range of masses can be obtained by considering boson stars with a self-interacting scalar field \citep{Grandclementetal14}.

Natural units of length and time are given by
\begin{equation}
\mathcal{L}=\frac{\hbar}{m\, c}
\end{equation}
and
\begin{equation}
\mathcal{T}=\frac{\hbar}{m\, c^2}.
\end{equation}
It is interesting to relate our unit of length to the usual unit of length in black-hole spacetime
\begin{equation}
\mathcal{L}'=\frac{G M}{c^2}.
\end{equation}
Let us consider a boson star with ADM mass $M=\mu \mathcal{M}$. It is immediate to show that
\begin{equation}
\mathcal{L}'=\mu \mathcal{L}.
\end{equation}
For all boson stars considered in this article, $\mu$ will be of order unity. As a consequence, the two units of length are comparable. In the following, all distances (in practice, all radii $r$) will be given in units of $\mathcal{L}'$, thus keeping the same units as in previous black-hole studies. This means in particular that the unit of length in this article is not the same as the one in \citet{Grandclementetal14}, which uses $\mathcal{L}$.

We also warn the reader accustomed to using the Kerr metric in Boyer-Lindquist coordinates that we will use here quasi-isotropic coordinates to keep the same coordinates for describing black holes and boson stars. As a consequence the coordinate radius $r$ used in this article for describing the Kerr metric is linked to the usual Boyer-Lindquist $r_{\rm BL}$ by
\begin{equation}
r = \frac{1}{2} \left(r_{\rm BL} - M + \sqrt{a^2 + r_{\rm BL} (r_{\rm BL} - 2 M)}\right),
\end{equation}
where $M$ and $a$ are the two standard parameters of the Kerr solution ($M$ is the ADM mass and $a$ the reduced spin).

\section{Circular orbits}

\subsection{Introduction}
Motivated by the construction of accretion tori, we start by analysing circular timelike geodesics, that will be called in the following circular orbits. They are described by their tangential 4-velocity $u^\mu=\left(u^t,0,0,u^\varphi\right)$ with the normalization condition $u^2=u_\mu\,u^\mu=-1$. We can deduce from the spacetime symmetries the existence of two constants of geodesic motion, namely the specific energy and angular momentum
\begin{eqnarray}\label{Eq_EL}
E&\,=\,&-u_{t}\,,\\
L&\,=\,&u_{\rm \varphi}.
\end{eqnarray}
Now we can introduce the functions that describe the rotation: the angular velocity $\Omega$ and the specific angular momentum $l$,
\begin{eqnarray}\label{Eq_Ol}
\Omega&\,=\,&\frac{u^\varphi}{u^{t}},\,\\
l&\,=\,&\frac{L}{E}\,=-\frac{u_\varphi}{u_{t}}.\,
\end{eqnarray}

The stability and general properties of circular orbits in the equatorial plane is of physical interest in the context of building accretion tori. We determine here the innermost circular orbit radius and study bound orbits \citep[following][]{Bardeenetal72}. The circular orbits satisfy two conditions, the effective potential $\mathcal{V}_{\rm eff}=0$ and its derivative $\partial \mathcal{V}_{\rm eff}/\partial r =0$. For a circular spacetime described by quasi-isotropic coordinates the effective potential in the radial direction is deduced from the equation governing the evolution of the radial position $r$ in the equatorial plane as a function of the particle's proper time $\tau$ \citep{Grandclementetal14, Gourg10} 
\begin{equation}\label{Eq_Veff0}
\left(\frac{\mathrm{d} r}{\mathrm{d} \tau}\right)^2-\mathcal{V}_{\rm eff}=0
\end{equation}
Thus, 

\begin{equation}\label{Eq_Veff}
\mathcal{V}_{\rm eff} = \frac{1}{A^2}\left(\frac{1}{\alpha^2}\,\left(E+\beta^{\varphi}\,L\right)^2-\frac{L^2}{B^2\,r^2}-1\right).
\end{equation}

We can deduce from this both the Keplerian angular momentum and energy
\begin{eqnarray}\label{Eq_lk}
l_{K\pm}\,&=&\frac{L_{K\pm}}{E_{K\pm}}\,=\,\frac{B\,r\,V_{\pm}}{\alpha-\beta^\varphi\,B\,r\,V_{\pm}}\,, \\
E_{K\pm}\, &=& \frac{\alpha - \beta^\varphi\,B\,r\,V_{\pm}}{\sqrt{1-V_{\pm}^2}}
\end{eqnarray}
with the circular orbit velocity $V_{\pm}$ with respect to the zero angular momentum observer (ZAMO) given by
\begin{equation}\label{Eq_l_K}
V_{\pm} = \frac{-\frac{B\,r}{\alpha}\frac{\partial \beta^\varphi}{\partial r}\pm\,\sqrt{D}}{2\,\left(\frac{\partial B}{B\partial r}+\frac{1}{r}\right)}\,
\end{equation}
and the $+$ and $-$ signs corresponding respectively to prograde and retrograde orbits \citep{Grandclementetal14}. Circular orbits exist when $D>0$, where
\begin{equation}\label{Eq_D_K}
D=\frac{B^2\,r^2}{\alpha^2}\left(\frac{\partial \beta^\varphi}{\partial r}\right)^2\,
+4\,\frac{\partial \ln\left(\alpha\right)}{\partial r}\,\left(\frac{\partial B}{B\partial r}+\frac{1}{r}\right)\,.
\end{equation}

\subsection{Existence of circular orbits}
Circular orbits exist if $r>r_{\rm ICO}$, where $r_{\rm ICO}$ is the radius of the innermost circular orbit. The value of $r_{\rm ICO}$ is a function of the quantum number $k$  and period number $\omega$ of the boson star. It increases with $k$ and $\omega$: we can check the variation of $r_{\rm ICO}$ in Table~\ref{Tab_KOMEGA}. It  corresponds to the  radius which admits a solution for the specific angular momentum.
The position of the ICO is always inside the torus of the boson star \citep{Grandclementetal14}.

For boson stars, all the circular orbits (i.e. $r>r_{\rm ICO}$) are found to be stable \citep{Grandclementetal14}, since the second derivative of the effective potential is always positive: $\partial^2 \mathcal{V}_{\rm eff} / \partial r^2 > 0$. This is why boson star spacetime have an ICO while Kerr spacetime has an ISCO (innermost \emph{stable} circular orbit), which is distinct from the ICO, since not all orbits are stable.

\subsection{Keplerian specific angular momentum}
\begin{figure*}
\resizebox{8cm}{4.8cm}{\includegraphics{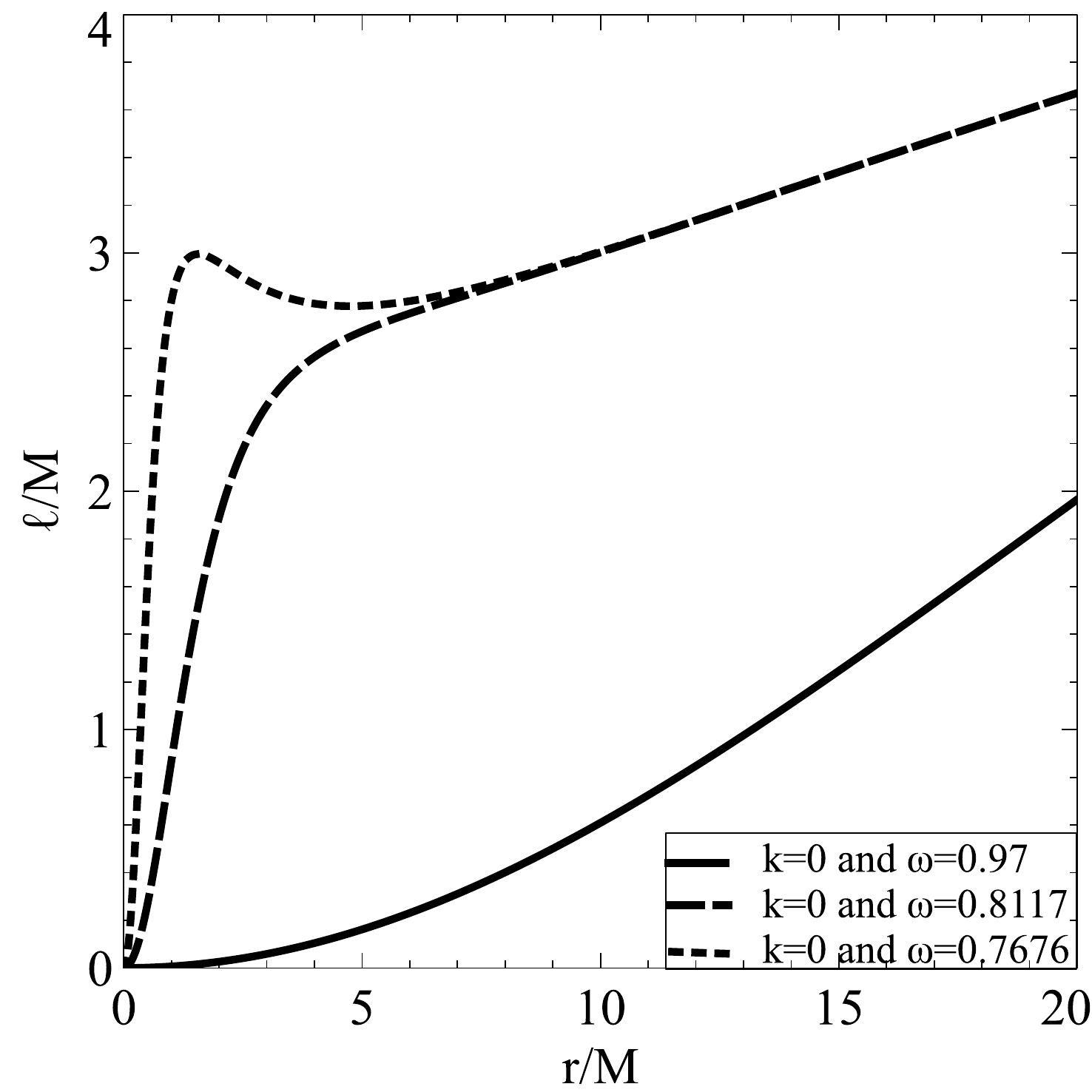}}
\resizebox{8cm}{4.8cm}{\includegraphics{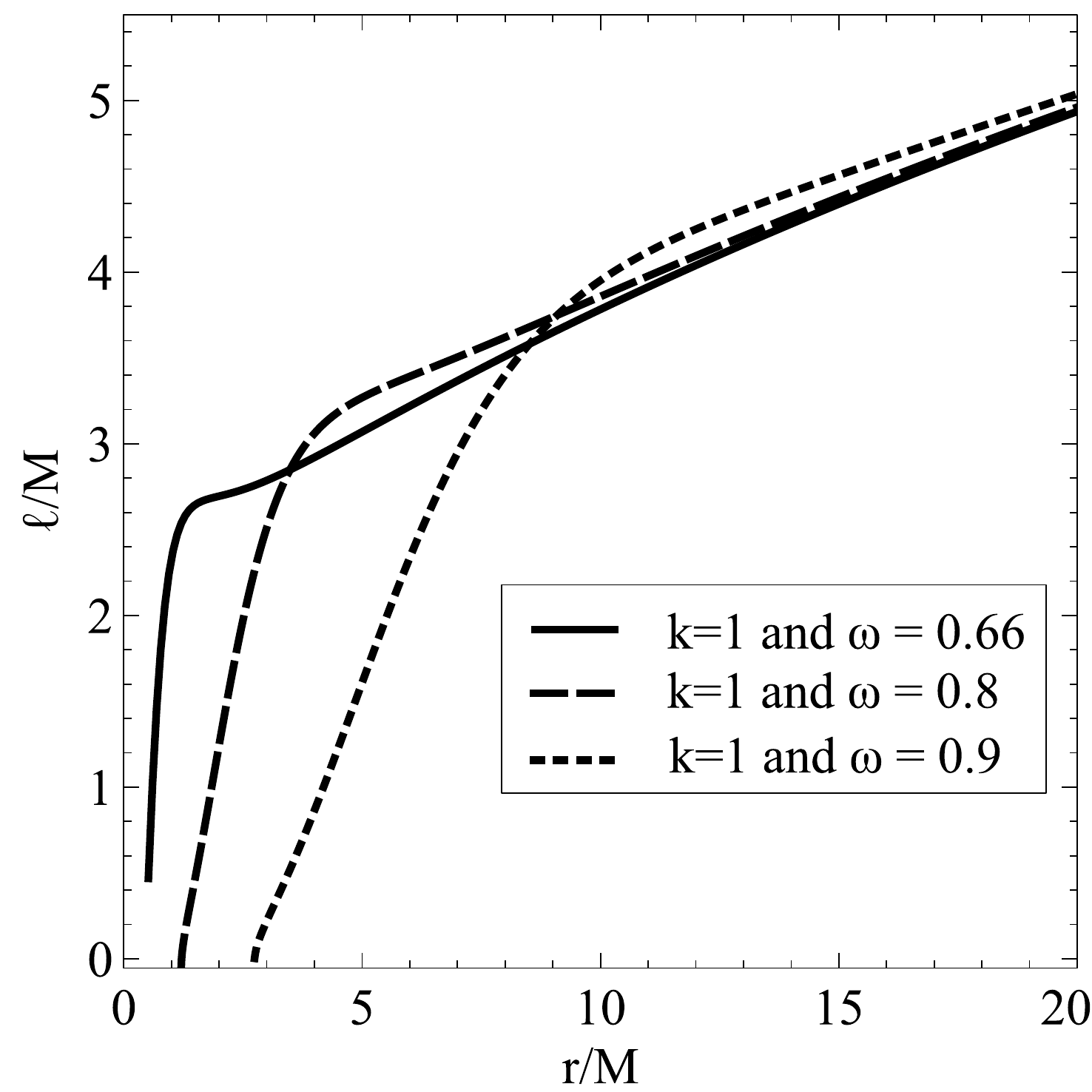}}
\resizebox{8cm}{4.8cm}{\includegraphics{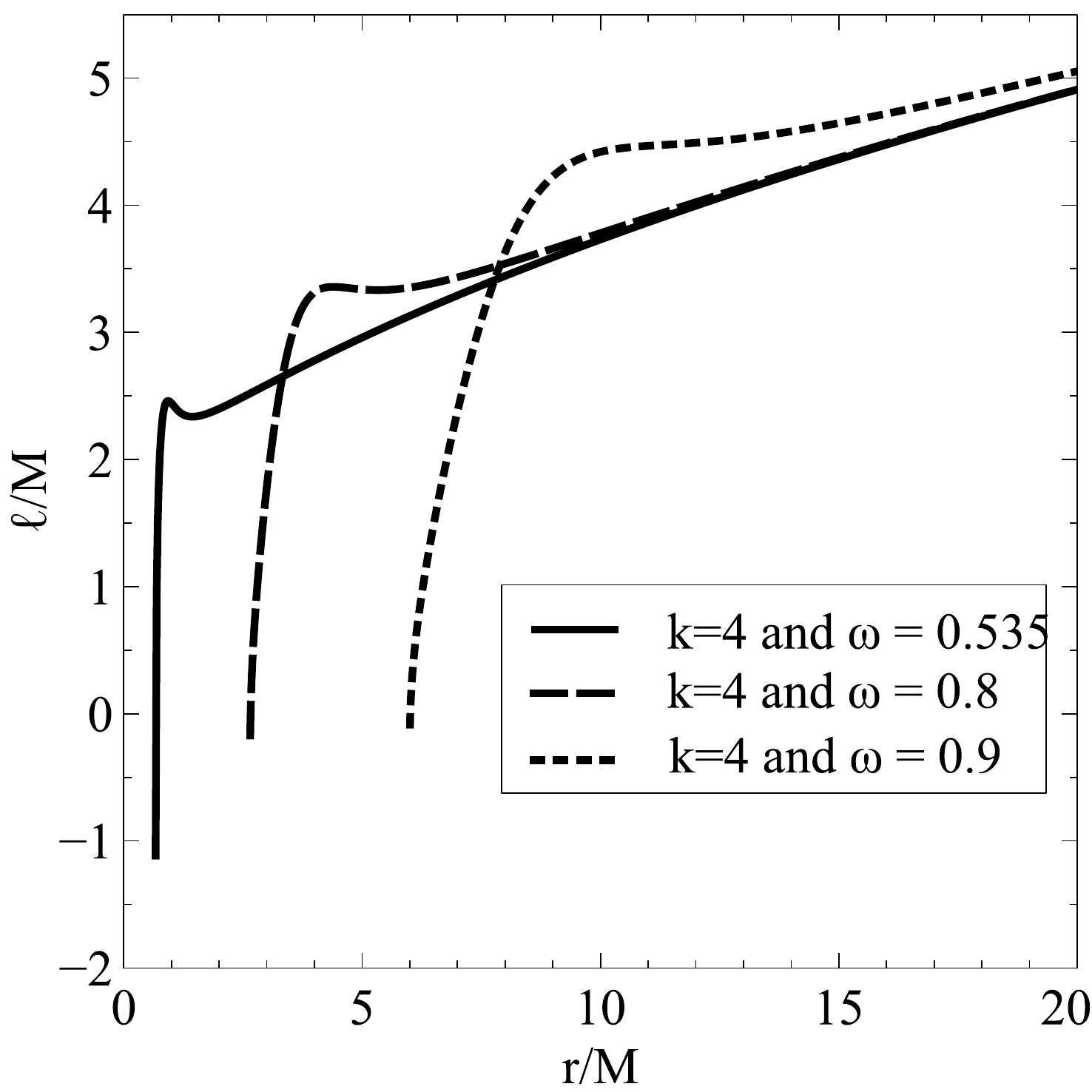}} 
\resizebox{8cm}{4.8cm}{\includegraphics{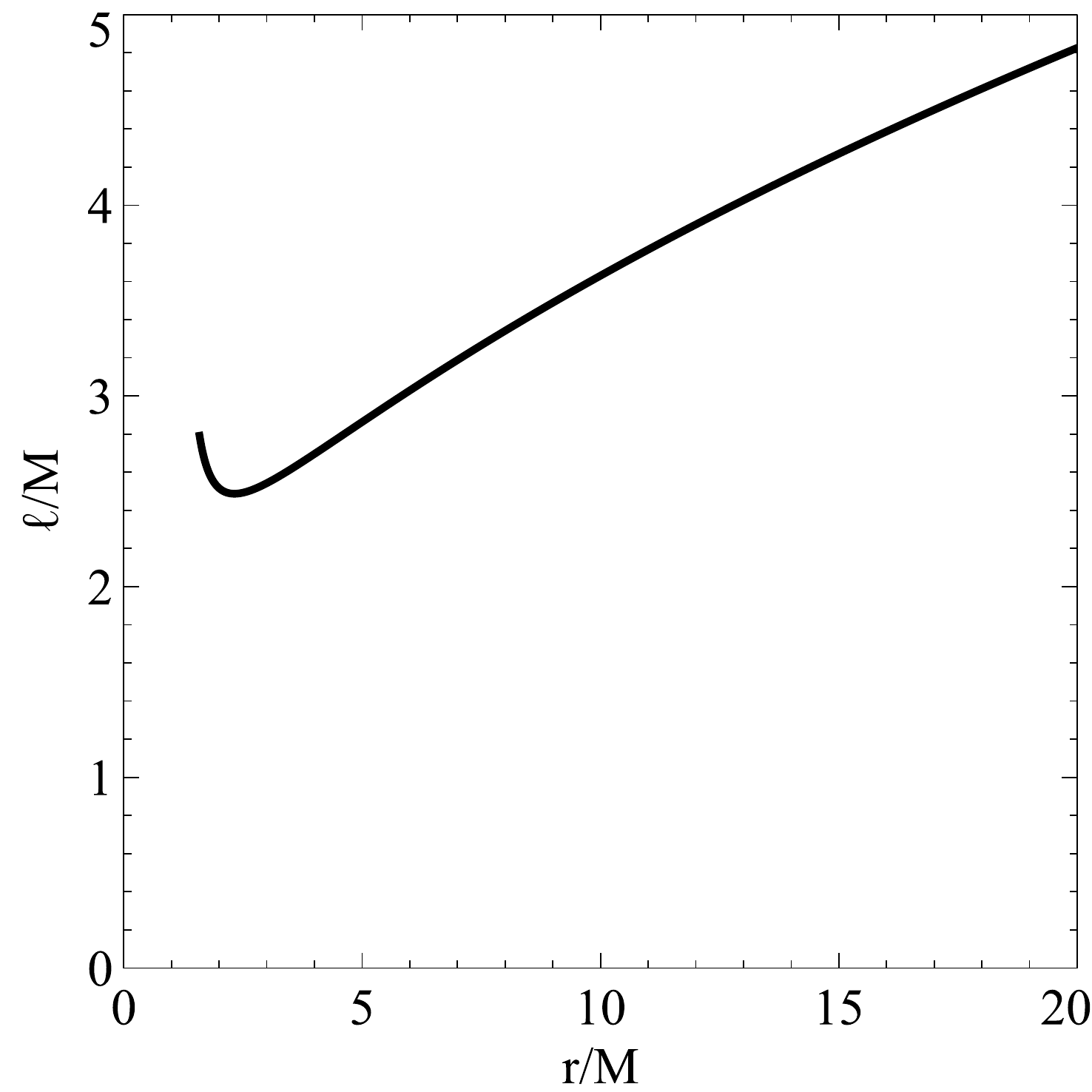}}
\caption{
Circular orbit specific angular momentum $l_{K+}$ given by Eq.~(\ref{Eq_lk}) for various boson stars (upper row, bottom left panel), together with the same quantity for the Kerr metric with $a/M=0.9$ (bottom right panel).
Boson star angular momenta are plotted for $r>r_{\rm ICO}$ while the Kerr angular momentum is plotted for all radii above the horizon. 
}

\label{Fig_allL}
\end{figure*}

Figure~\ref{Fig_allL} shows Keplerian angular momentum profiles for various boson star spacetimes and for a fast-rotating Kerr black hole $a=0.9$ for comparison. All boson stars converge to the same value of angular momentum as in Kerr at large radii, where they all follow a $r^{1/2}$ profile. It is of course in the strong-field region where different behaviours are observed. The well-known Kerr case shows a minimum (at the ISCO) followed by an increase of the angular momentum with decreasing radius until the event horizon is reached. On the contrary for boson stars, the angular momentum is either a monotonically decreasing function of  $r$ with a value close to $0$ (could become negative) at the ICO, or a more complex profile with a local maximum close to the ICO. This second case is obtained for the most relativistic (i.e. small $\omega$) boson stars and the limiting value of $\omega$ increases for faster rotating stars.

This ``relativistic profile'' of the angular momentum is particularly interesting in the prospect of building accretion tori such as Polish doughnuts with constant angular momentum. It is well-known that Polish doughnuts surrounding black holes have a centre (the pressure maximum) where the fluid has exactly Keplerian rotation and may also have a ``cusp'', i.e. a point where the equipotential surface crosses itself, from which matter can flow onto the black hole and where the rotation is also Keplerian \citep[see e.g.][]{abramowicz09}. Such a configuration is possible in Kerr geometry because the angular momentum profile has a minimum (bottom right panel of Fig.~\ref{Fig_allL}) which allows a constant line to intersect with  the Keplerian profile at two radii.  The boson star spacetime, however, allows for more than one points of intersection only in the sufficiently relativistic  cases where an extremum of the Keplerian angular momentum profile exists. We study this feature in more detail in Sec.~\ref{sec:tori}.

We have mentioned above that the Keplerian angular momentum reaches values close to zero at the ICO. Actually, it can even become slightly negative,  for instance in the case where $k=4$ and $\omega=0.535\; m/\hbar$, shown in the lower left panel of Fig.~\ref{Fig_allL}. This behaviour is common to all rotating boson stars even at smaller $k$ and bigger $\omega$ and is linked to the sign of the prograde Keplerian velocity $V_{K+}$. This quantity has the same sign as its numerator, $-B\,r / \alpha \,\partial \beta^\varphi/\partial r + \sqrt{D}$. It can become negative if the second term  under the square root in the definition of $D$ becomes negative, $\partial \ln\alpha / \partial r \,\left(1/B\,\partial B/ \partial r+1/r\right)<0$. This condition is verified for rotating boson stars very slightly above the ICO, leading to the change of sign of the prograde Keplerian velocity and hence of the angular momentum $l_{K+}$. We note that although a particle on a prograde orbit will appear to rotate backwards to the ZAMO, it will still have a positive angular velocity as observed from infinity.

\subsection{All circular orbits are bound}
All the circular orbits with $r>r_{\rm ICO}$ have  $E=\gamma\,\left(\alpha-\beta^\varphi\,B\,r\,V\right) \leq 1$ and thus they are bound.

\begin{figure*}
\resizebox{8cm}{4.8cm}{\includegraphics{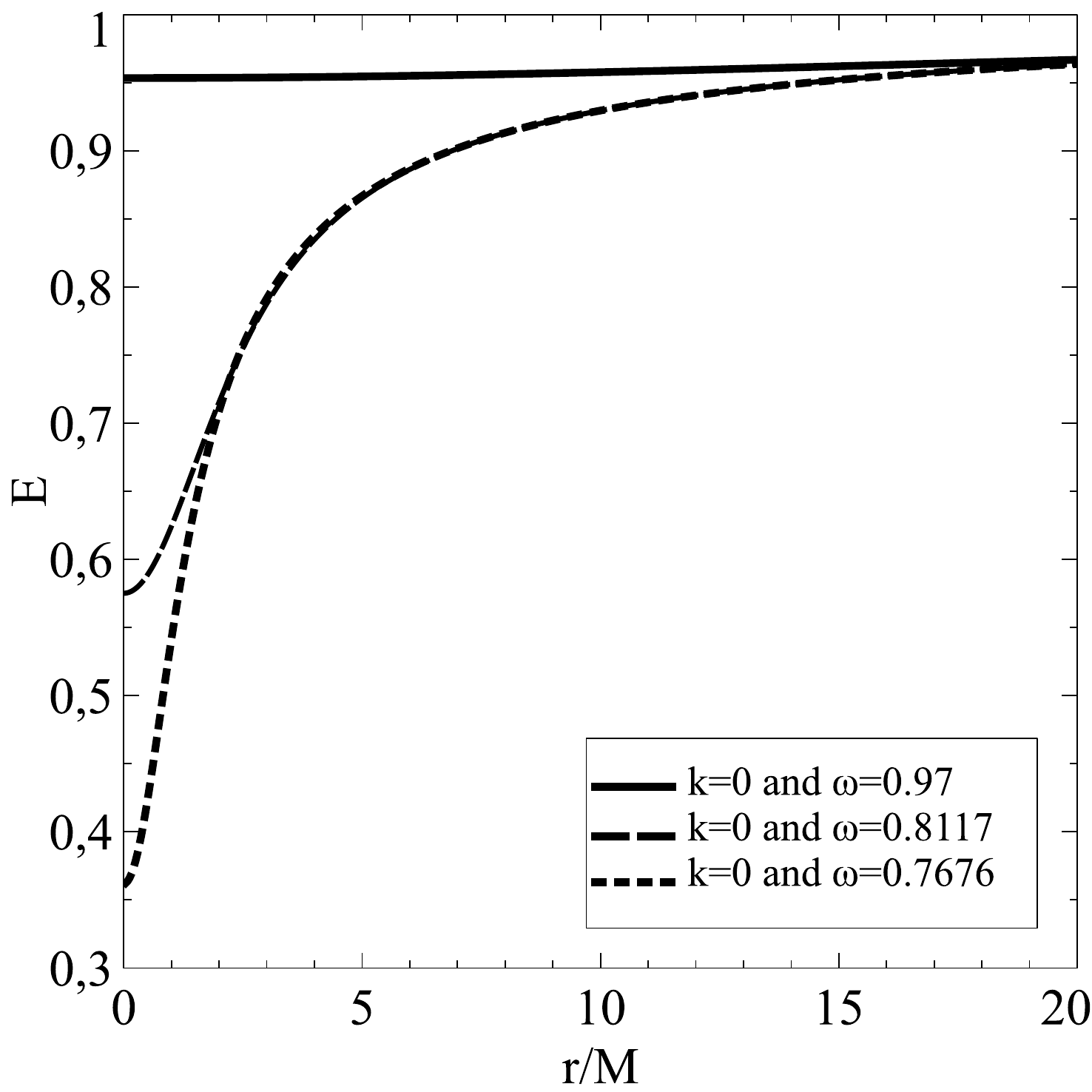}}
\resizebox{8cm}{4.8cm}{\includegraphics{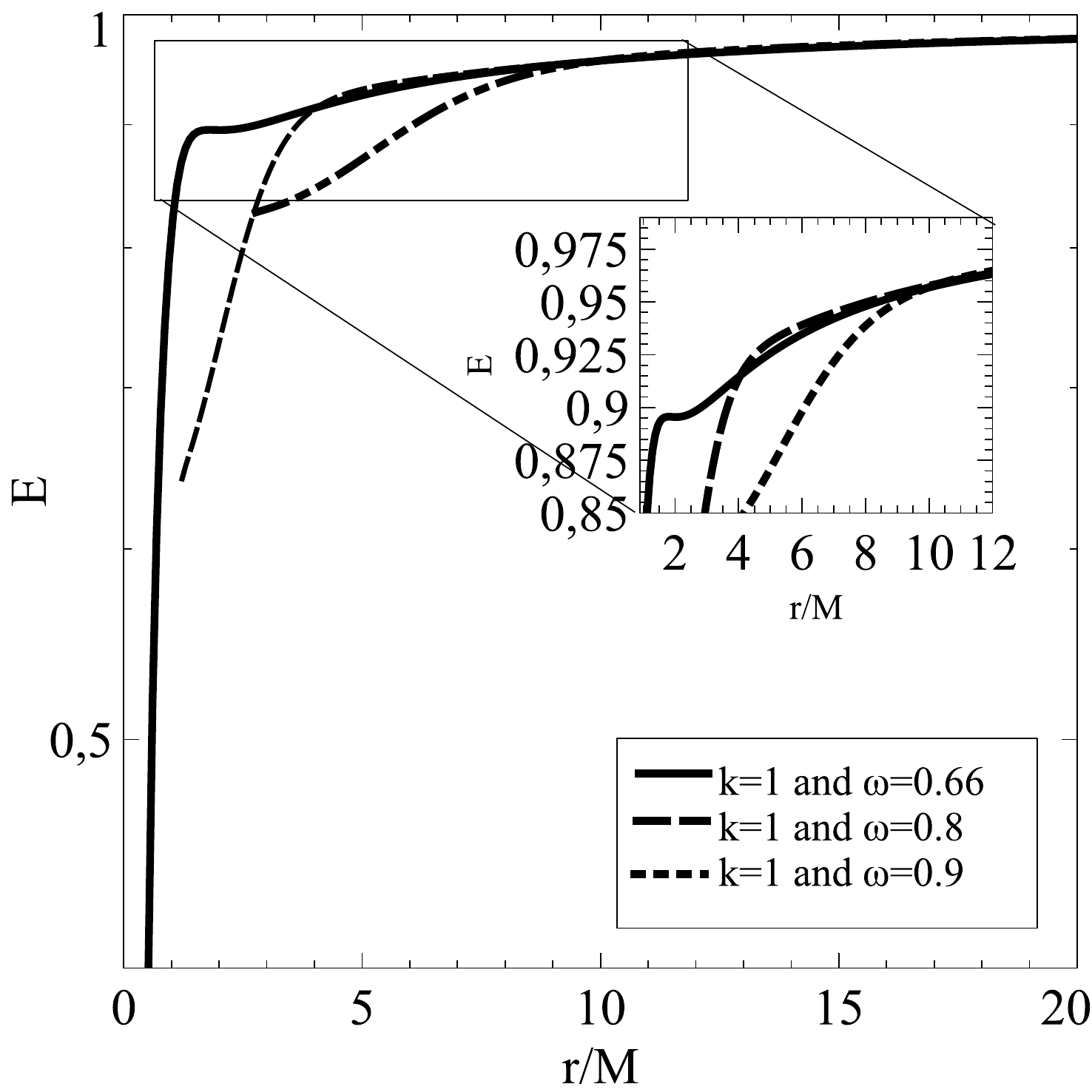}}
\resizebox{8cm}{4.8cm}{\includegraphics{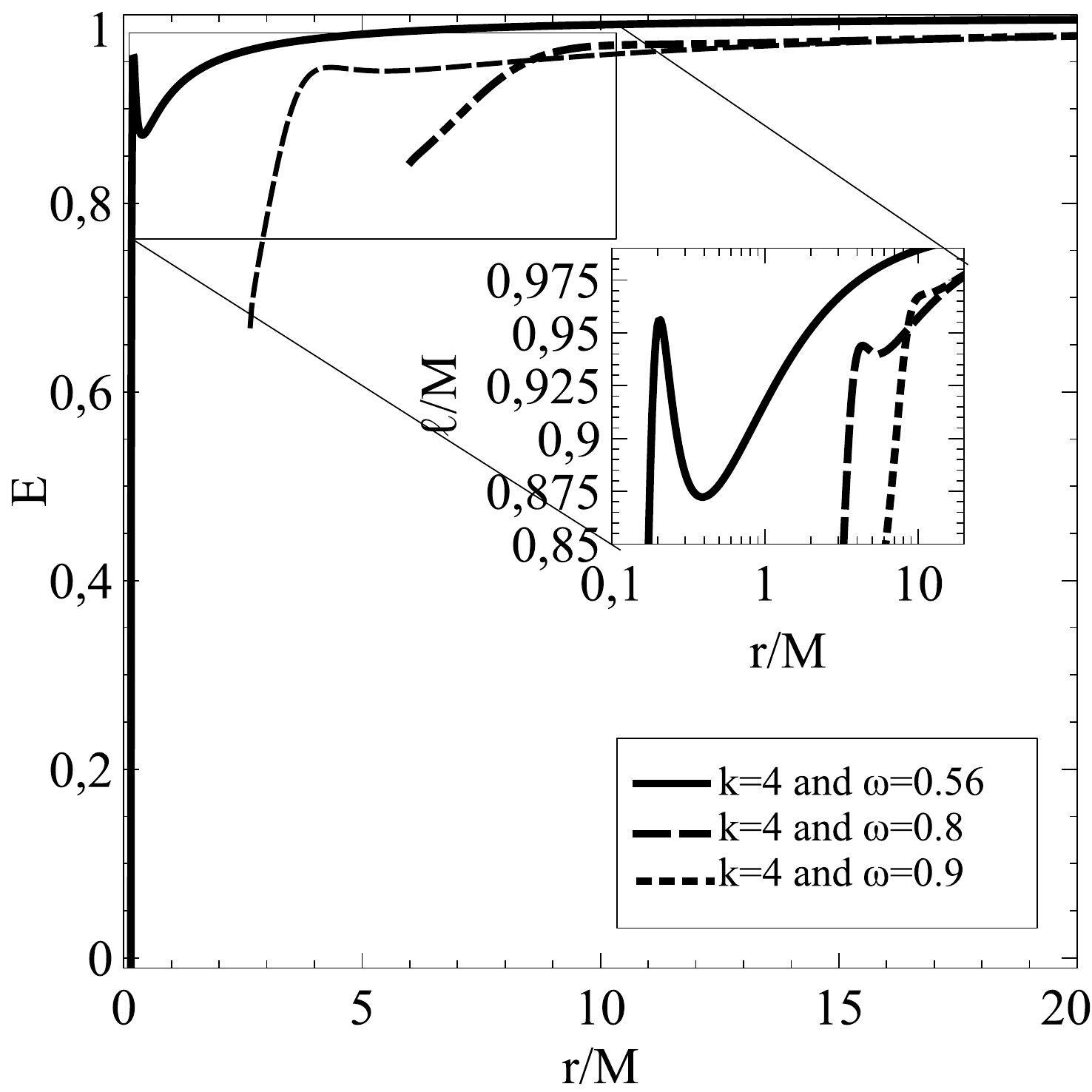}} 
\resizebox{8cm}{4.8cm}{\includegraphics{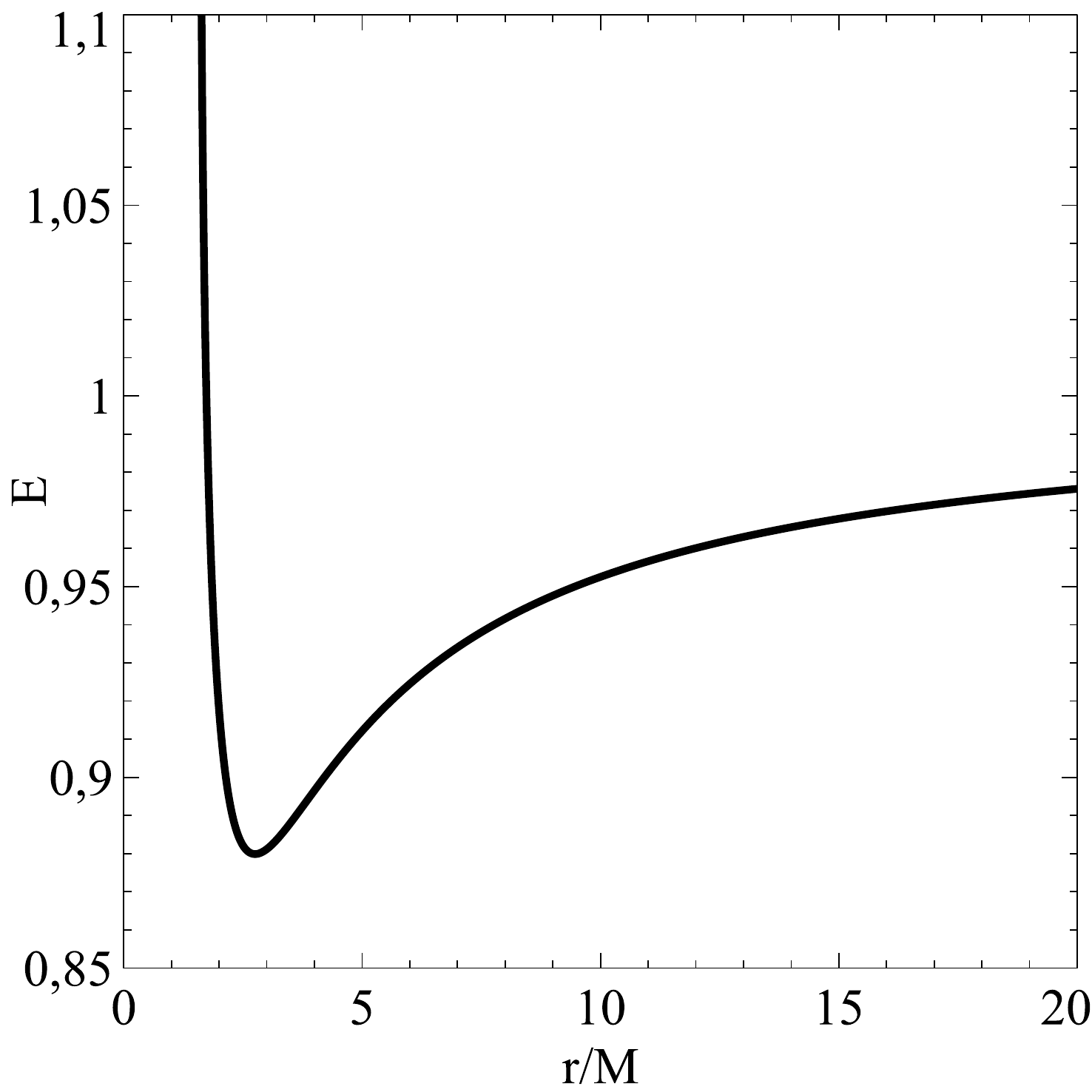}}
\caption{Same as Fig.~\ref{Fig_allL} for the specific energy $E$ of circular
orbits.}
\label{Fig_Ekall}
\end{figure*}

When we compare the circular orbit energy between the boson star case and the Kerr black hole case, the difference appears at $r<r_{0}$ (see Fig.~\ref{Fig_Ekall}), where $r_{0}$ is the radius where the energy changes its slope. In the case of Kerr black holes, the energy starts to increase going towards smaller $r$ and becomes higher than one (Fig.~\ref{Fig_Ekall}, lower right panel), meaning the orbits become unbound. The bosons stars with $\omega$ close to $m/\hbar$ (i.e. the less relativistic ones), show an energy that decreases with the radius and that remain at energies $E<1$. The highly relativistic boson stars are different in the sense that the energy admits a maximum before decreasing close to the centre. Nevertheless, this maximum is always below $1$ and thus all the circular orbits are bound. From  Table~\ref{Tab_KOMEGA} we can see that even for the slowly rotating boson stars with $a/M<1$, circular orbits are always bound unlike for Kerr black holes.

\subsection{Spherical boson stars, $k=0$}
In terms of orbits, the boson stars behave in a very similar manner to their counterparts with $k>0$: all the orbits are stable and bound. The main difference is that there is no ICO, i.e. circular orbits exists down to the centre of the boson star, $r=0$ (see Fig.~\ref{Fig_LlS}). For the less relativistic star considered (corresponding to $\omega=0.9762\; m/\hbar$), the circular orbit has an almost constant energy with distance and thus has a similar behaviour as in the classical case (Fig.\ref{Fig_Ekall}, Top).

 \section{Matter tori with constant angular momentum}\label{sec:tori}

\subsection{Building Polish doughnuts}
The model of stationary relativistic thick matter tori with constant angular momentum is well  established. It was first derived by \cite{FishboneetMoncrief76} for an isentropic torus and later by \cite{Kozlowskietal78} for a torus with a barotropic equation of state (EOS). In the following, we  present the relevant aspects needed to build a thick torus in the case of a barotropic EOS. With this EOS, the surfaces of constant angular momentum $l$ coincide with the surfaces of constant angular velocity $\Omega$ and are called von Zeipel cylinders. In the case of purely circular motion with constant specific angular momentum it can be shown that for a stationary axisymmetric configuration the relativistic Euler equation admits the following first integral (\citet{Abramowiczetal78}; see also Sec.~2.2 of \citet{Straubetal12}):
\begin{equation}\label{e:first_integral}
H-H_{\rm in} = - \ln\left(-u_t\right) + \ln\left(-u_{t, \rm in}\right),
\end{equation}
where $H$ is the enthalpy function defined by
\begin{equation}\label{e:def_H}
H = \int^P_0 \frac{{\rm d}P}{\rho(1+\epsilon) + P}
\end{equation}
and the index `in' stands for values at the inner radius $r_{\rm in}$ of the torus. The resulting torus is known as a Polish doughnut and is characterized by the potential $W=\ln\left(-u_{t}\right)$, which, as a consequence of (\ref{e:first_integral}) and (\ref{e:def_H}), satisfies
\begin{equation}\label{e:gradP_gradW}
\frac{\nabla_{\mu}P}{P+\rho(1+\epsilon)} = - \nabla_{\mu} W.
\end{equation}

In the case of a torus with a constant angular momentum, the fluid 4-velocity can be easily derived. It satisfies in particular
\begin{equation}
u_t^2=\frac{\alpha^2 B^2 r^2\sin^2\theta}
{
\left(-\alpha^2+{\beta^{\varphi}}^2 B^2 r^2\sin^2\theta
\right) l^2+2 \beta^{\varphi}\,B^2 r^2
l \sin\theta +B^2 r^2
\sin^2\theta\
}.
\label{eq_ut_constL}
\end{equation}
The numerator is always strictly positive for $r>0$ (it is equal the radial Weyl coordinate squared $\rho^2=g_{t\varphi}^2-g_{tt}g_{\varphi\varphi}$ which is always positive outside the horizon in the Kerr spacetime). However, the denominator can be positive or negative. The Polish doughnut is of course only defined when the denominator is strictly positive. Solving for $l$, the zero of the dominator leads to two solutions

%
\begin{eqnarray}\label{eq_limit_L}
l_{\rm max} = \frac{\beta^{\varphi}\,B^2 r^2+\sqrt{\delta}}{\left(\alpha^2-{\beta^{\varphi}}^2 B^2 r^2
\right)}\,\\
l_{\rm min} = \frac{\beta^{\varphi}\,B^2 r^2-\sqrt{\delta}}{\left(\alpha^2-{\beta^{\varphi}}^2 B^2 r^2\right)}
\end{eqnarray}
where $\delta=\left(\beta^{\varphi}\,B^2 r^2\right)^2-\left(-\alpha^2+{\beta^{\varphi}}^2 B^2 r^2
\right)\,B^2 r^2$, we have checked that outside the ergosphere the denominator is positive if the angular momentum is between $l_{\rm min}$ and $l_{\rm max}$, while inside the ergosphere it is positive if the angular momentum is smaller than $l_{\rm max}$ or larger than $l_{\rm min}$ (in the ergosphere $l_{\rm min} > l_{\rm max}$). The quantity $l_{\rm min}$ is negative outside the ergosphere and becomes positive only inside.

\begin{figure}
\resizebox{8cm}{4.8cm}{\includegraphics{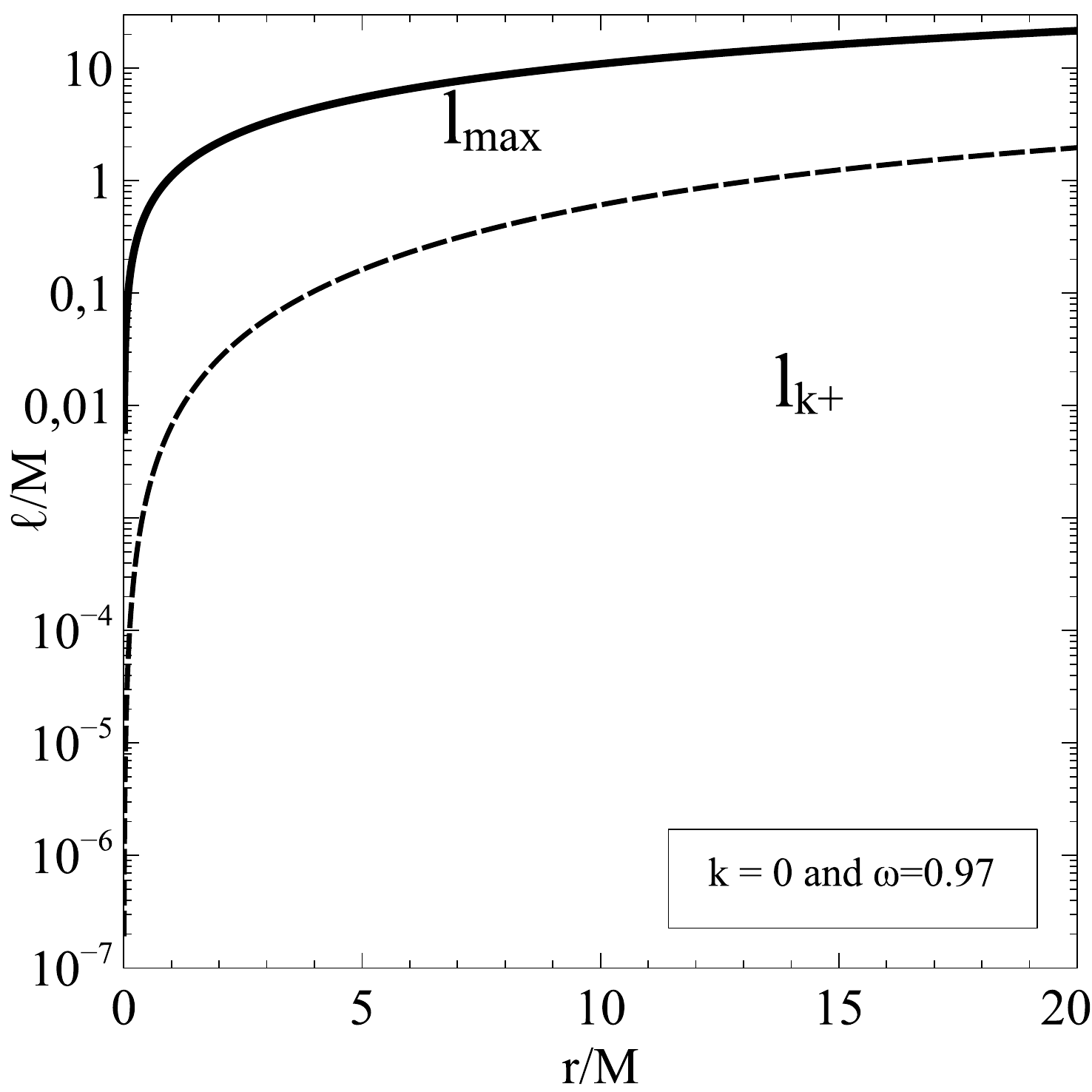}}
\resizebox{8cm}{4.8cm}{\includegraphics{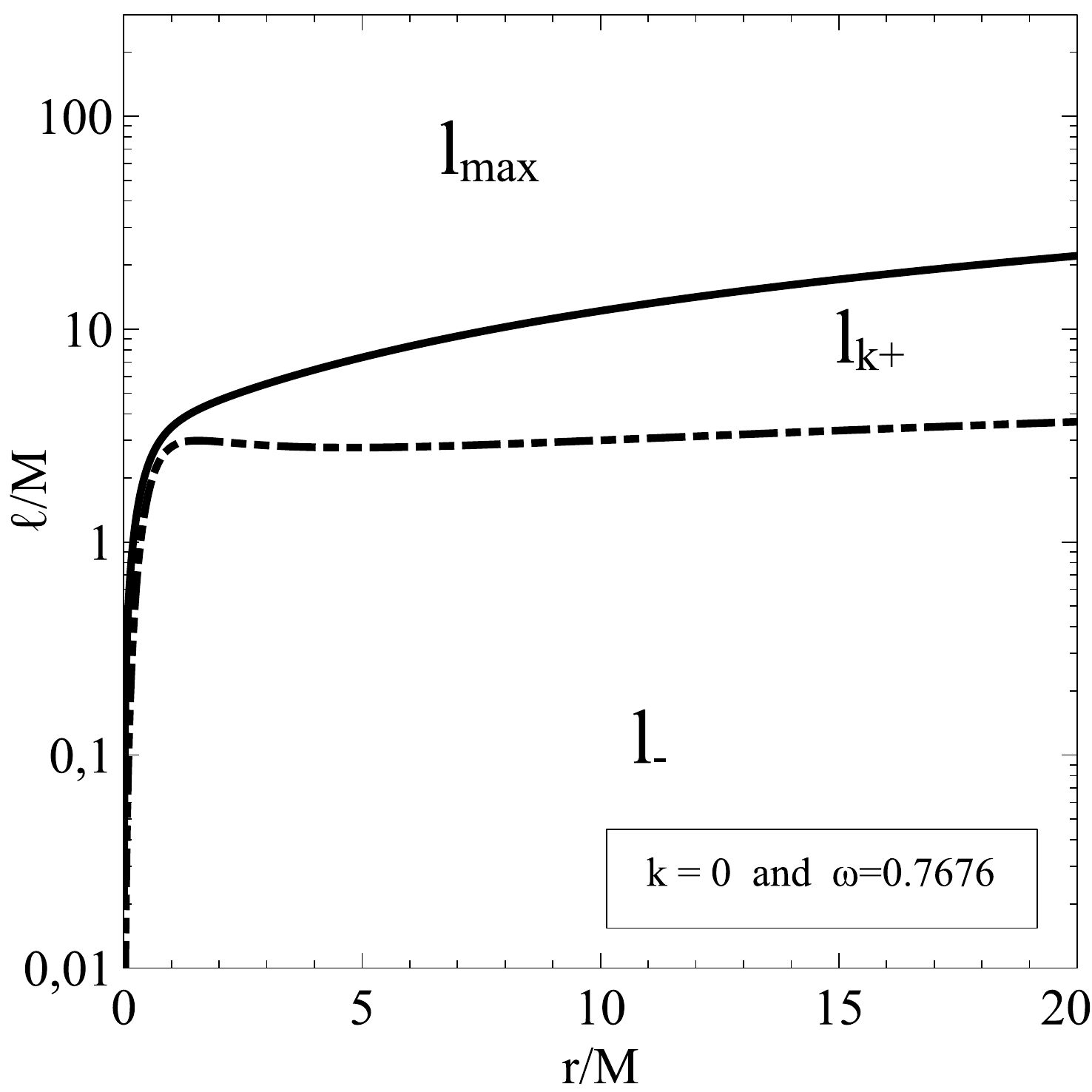}}
\caption{
The maximum allowed value of specific angular momentum as a function of the radius $r$ for spherical boson stars. The top figure is for a boson star with $\omega=0.97\; m/\hbar$, while the bottom one is for  $\omega=0.7676\; m/\hbar$.
}
\label{Fig_LlS}
\end{figure}

\begin{figure*}
\resizebox{8cm}{4.8cm}{\includegraphics{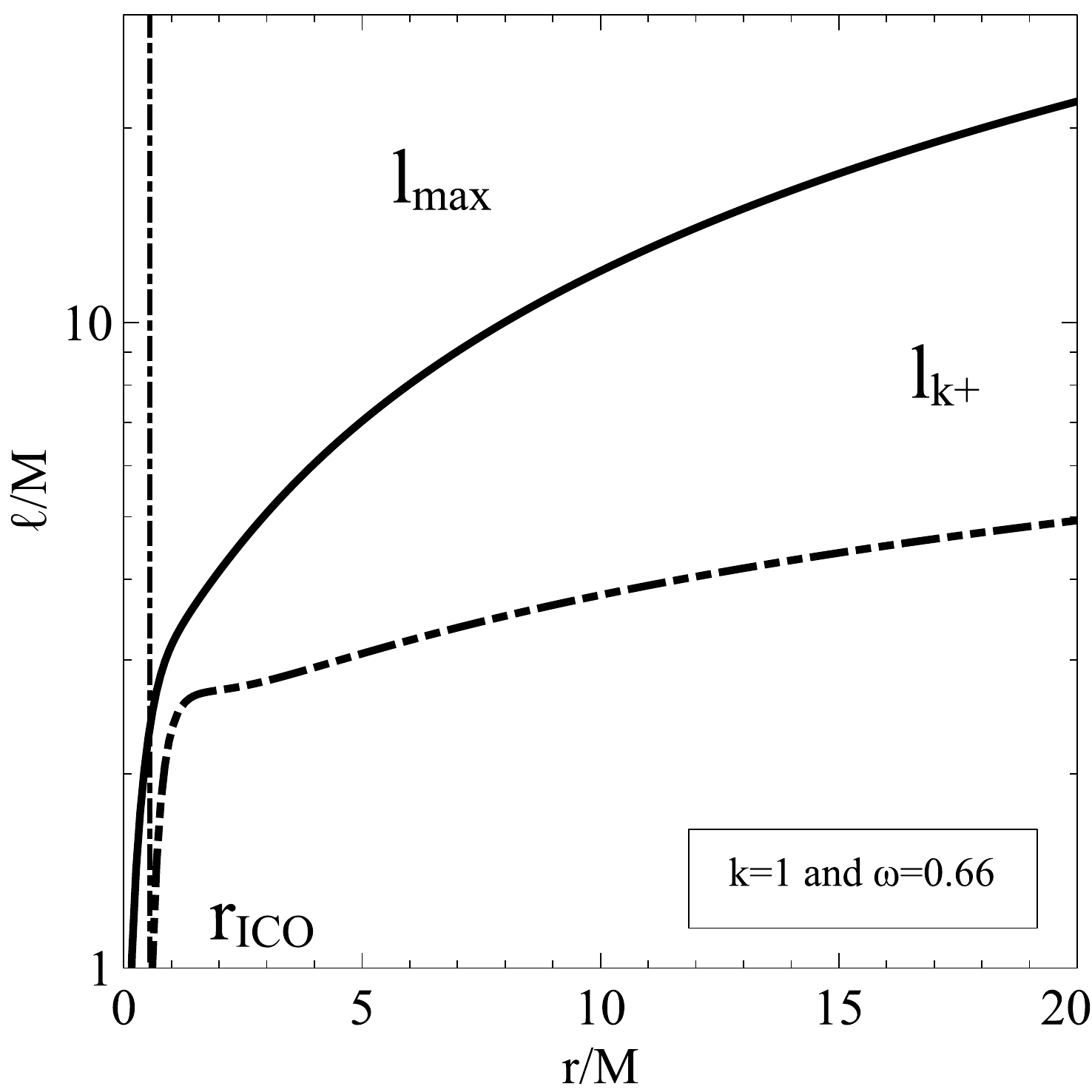}}
\resizebox{8cm}{4.8cm}{\includegraphics{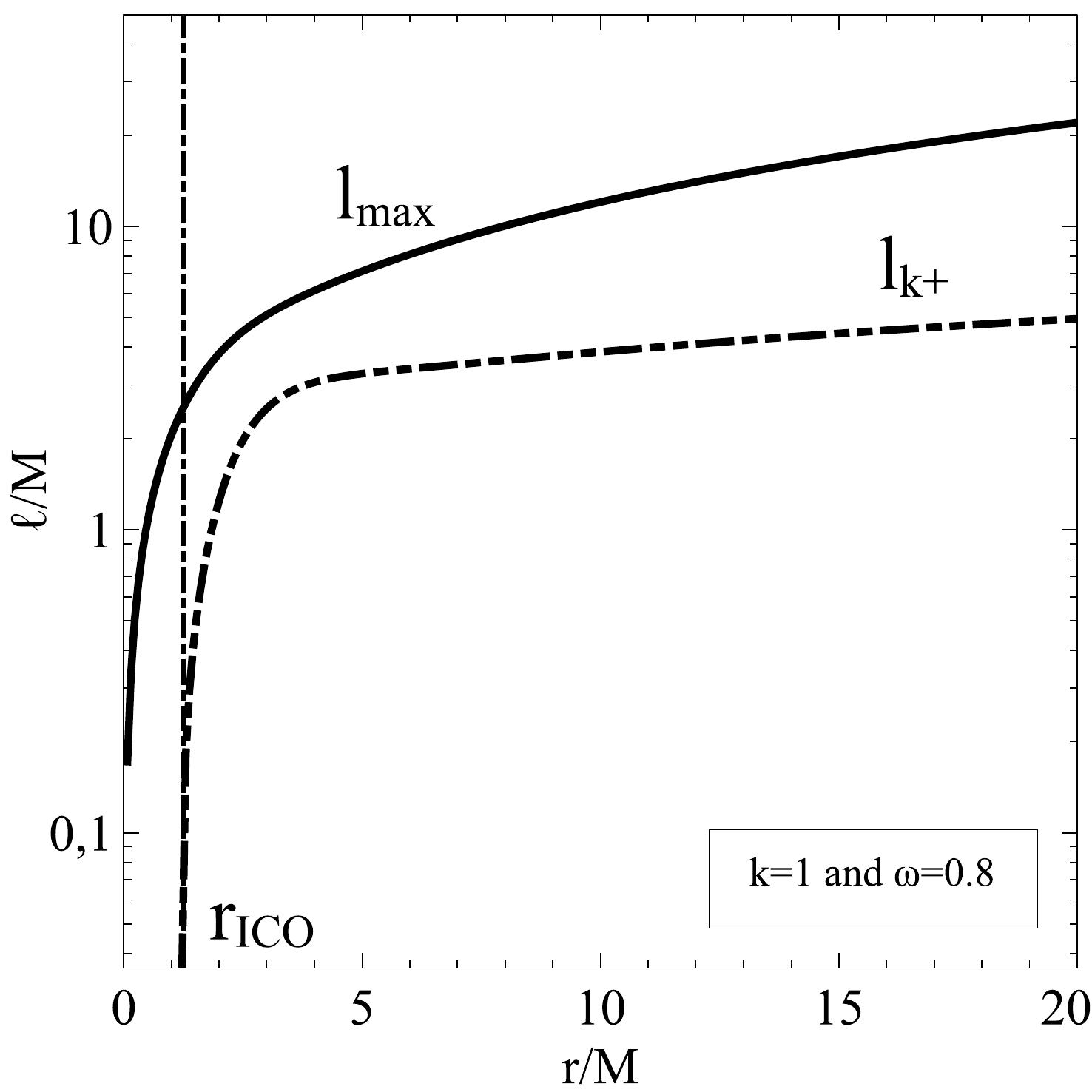}}
\resizebox{8cm}{4.8cm}{\includegraphics{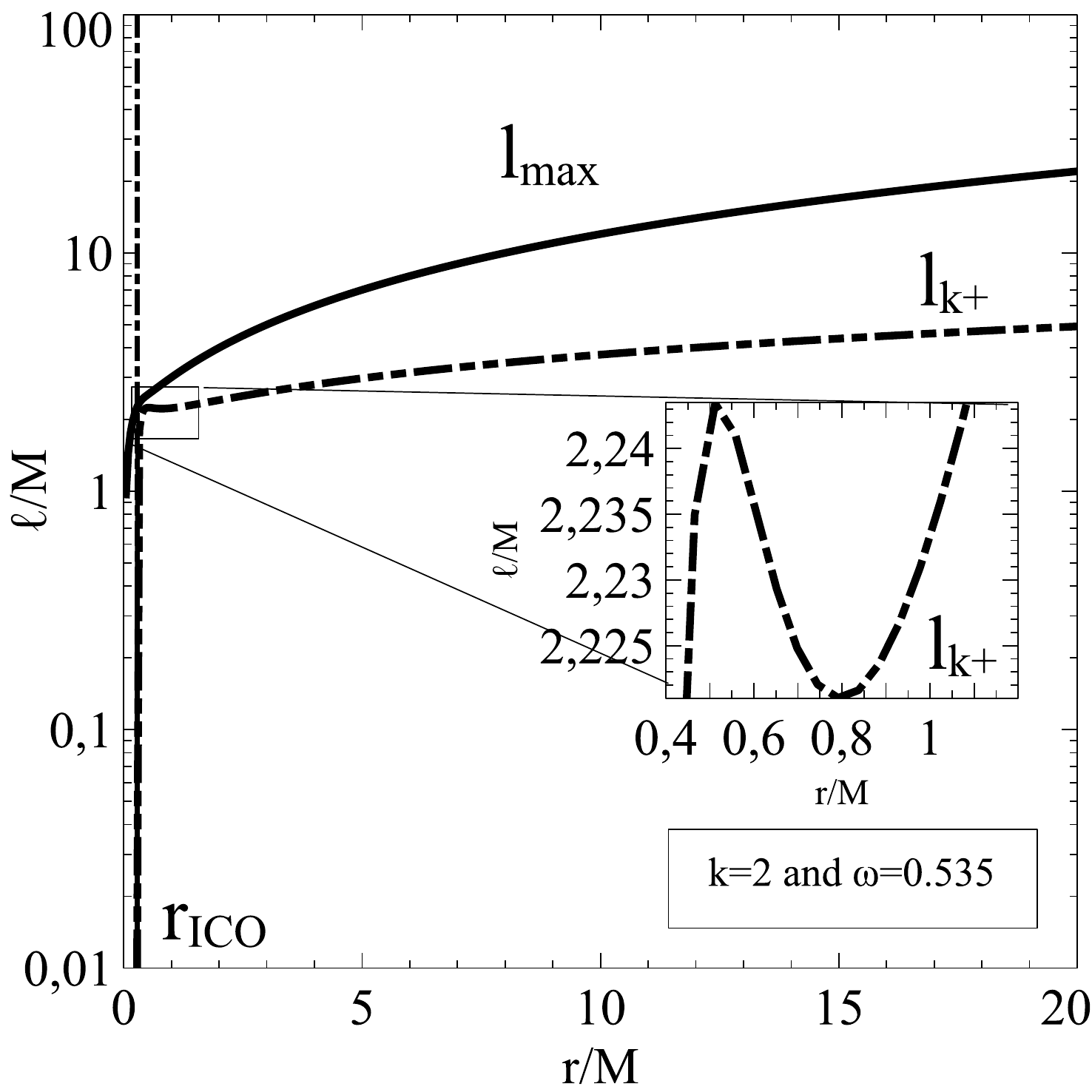}}
\resizebox{8cm}{4.8cm}{\includegraphics{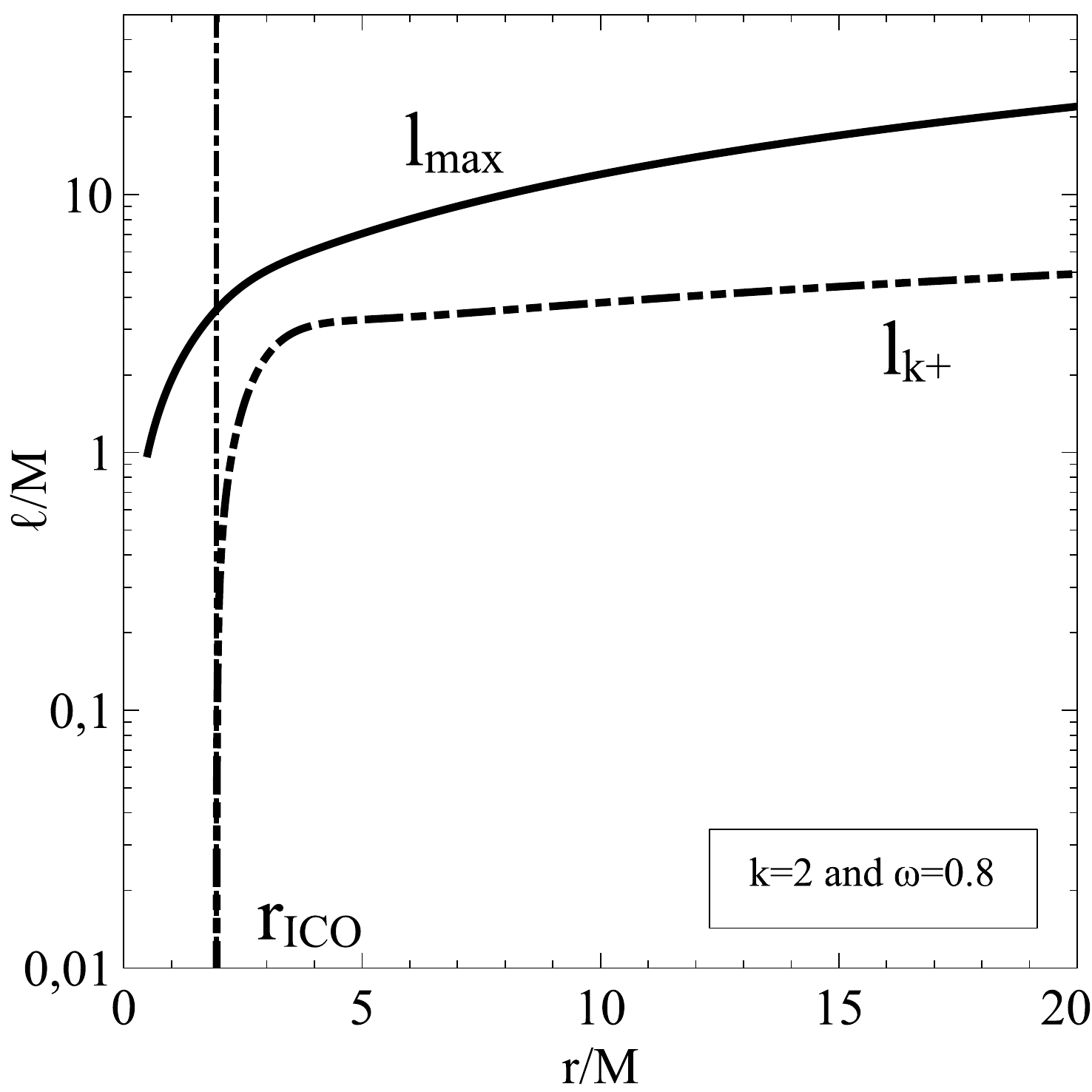}}
\resizebox{8cm}{4.8cm}{\includegraphics{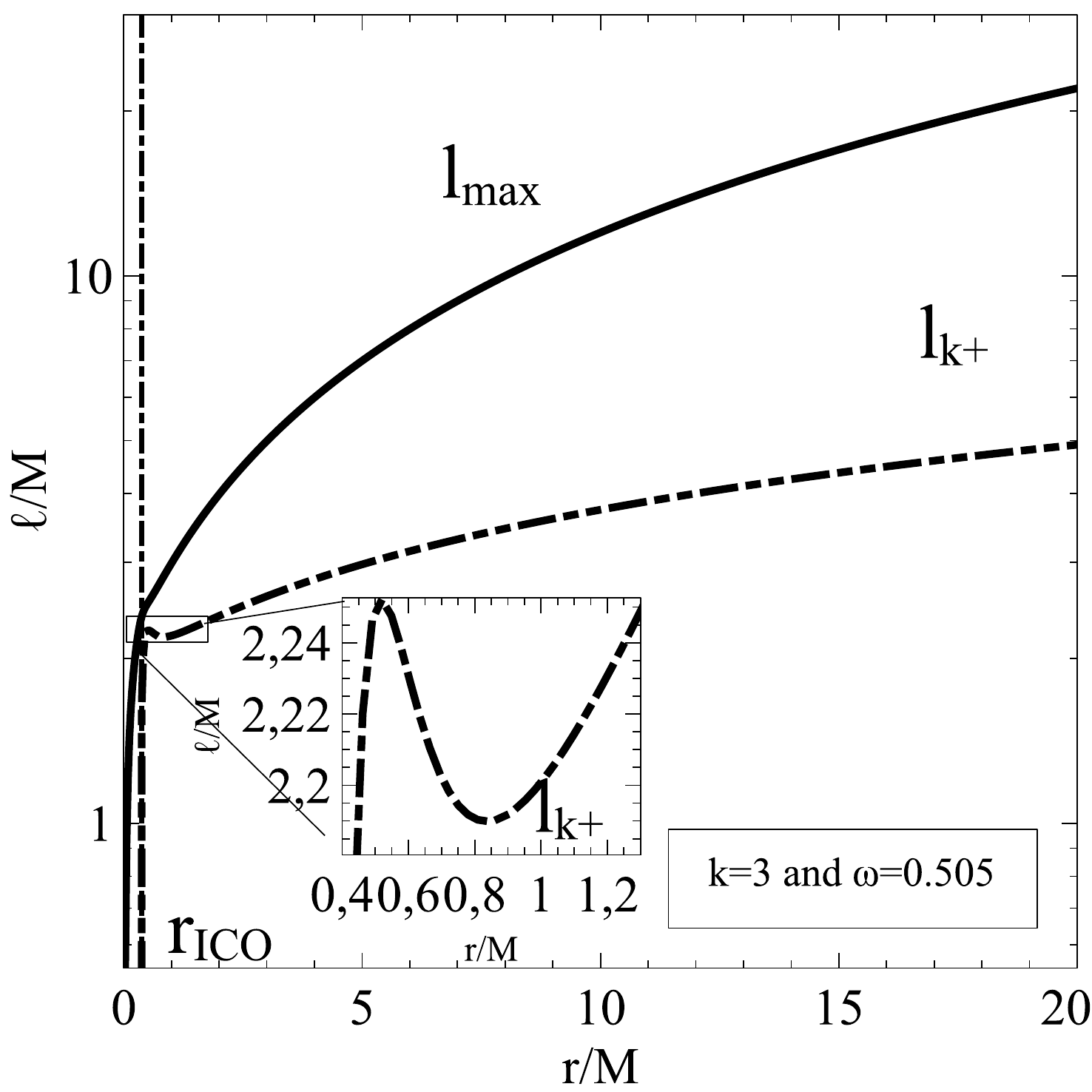}}
\resizebox{8cm}{4.8cm}{\includegraphics{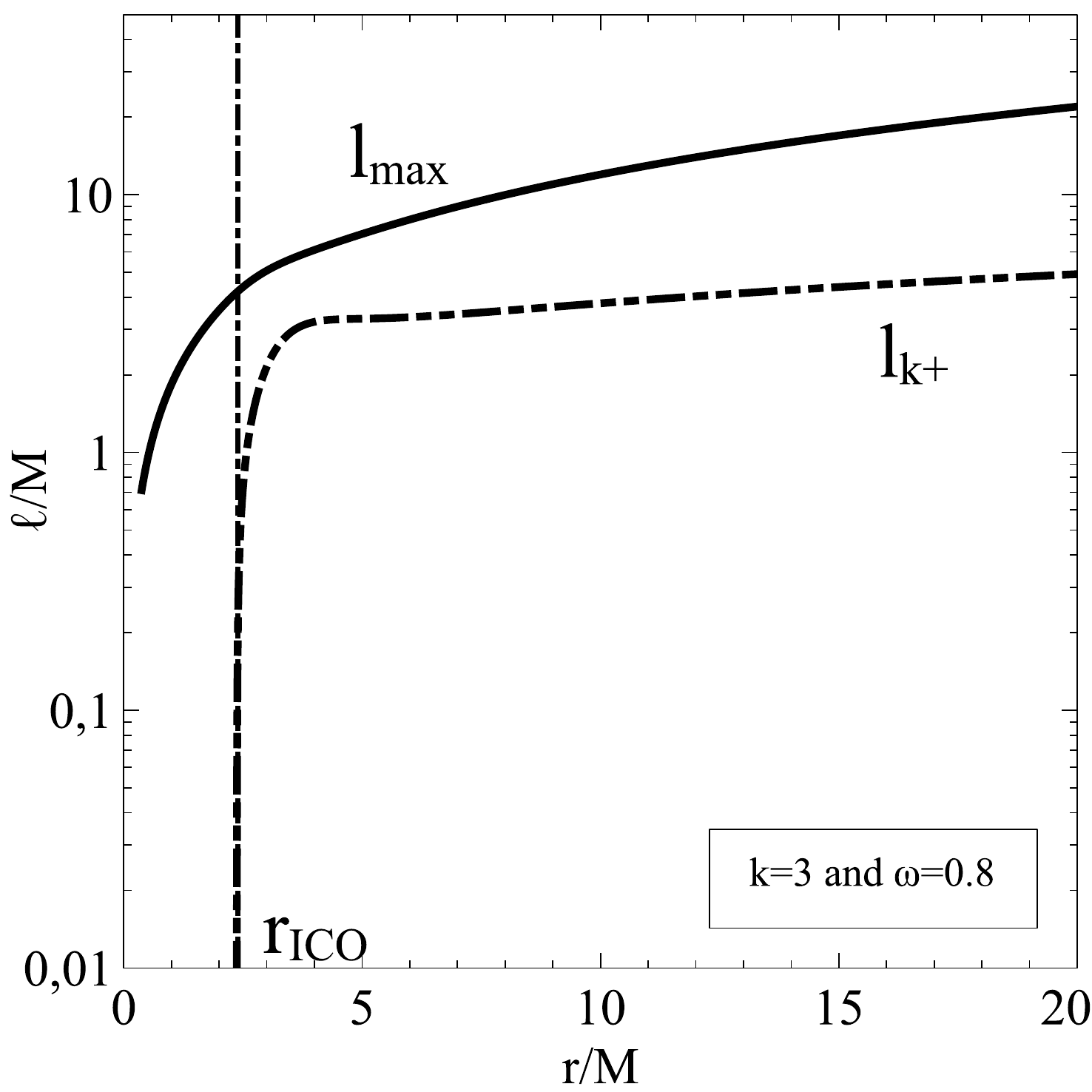}}
\resizebox{8cm}{4.8cm}{\includegraphics{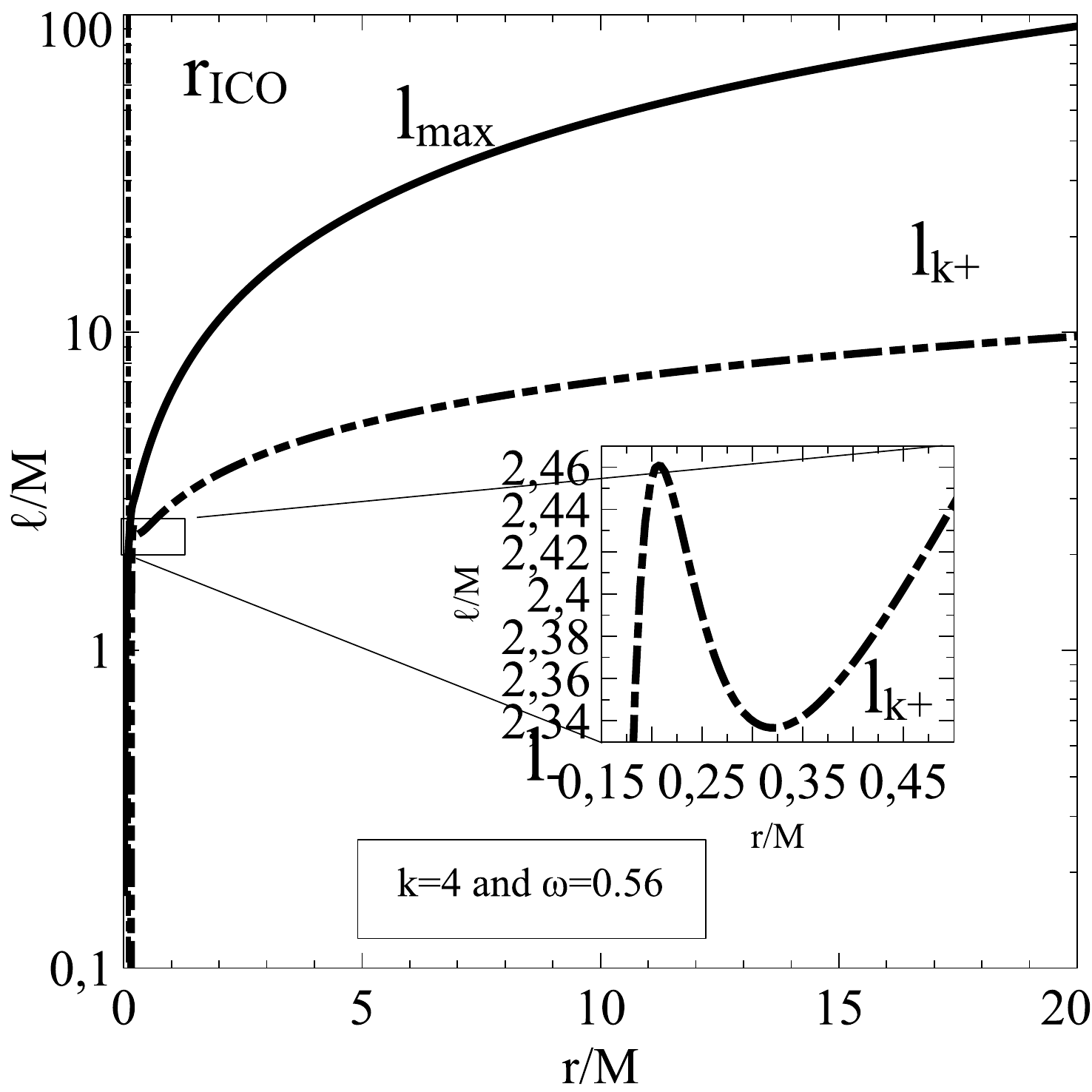}}
\resizebox{8cm}{4.8cm}{\includegraphics{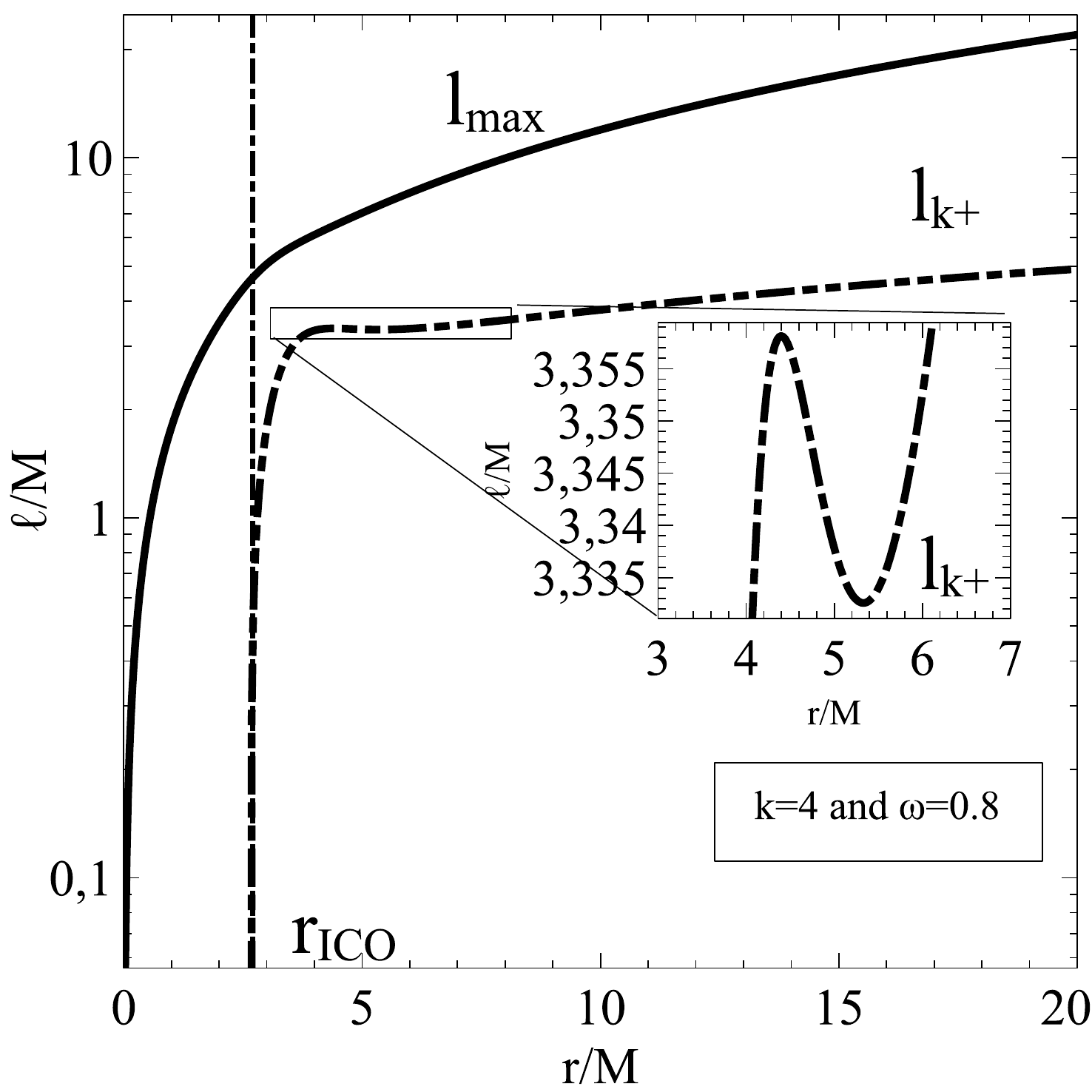}}
\caption{
The maximum value of the allowed specific angular momentum for torus with constant angular momentum $l_{\rm max}$  and  the Keplerian specific angular momentum $l_{\rm k}$.
}
\label{fig_Llimitut}
\end{figure*}

For a barotropic EOS, rest-mass density and pressure are related through $P=\kappa \rho^\Gamma$ where $\kappa$ is the polytropic constant and $\Gamma$ the polytropic index. They have the following expressions inside the Polish doughnut

\begin{eqnarray}
\rho&\,=\,&\left(\frac{\Gamma-1}{\Gamma}\frac{\frac{u_{\rm t, in}}{u_{t}}-1}{\kappa}\right)^{\frac{1}{\Gamma-1}}\,,\\
P&\,=\,&\kappa\,\left(\frac{\Gamma-1}{\Gamma}\frac{\frac{u_{\rm t, in}}{u_{t}}-1}{\kappa}\right)^{\frac{\Gamma}{\Gamma-1}}.
\end{eqnarray}

Let $r_{\rm centre}$ be the value of $r$ at which $\nabla_\mu W=0$. Thanks to Eq.~(\ref{e:gradP_gradW}), the pressure gradient vanishes as well at $r_{\rm centre}$ and the Keplerian specific angular momentum at $r=r_{\rm centre}$ is equal to the constant $l$. At this position, the density and pressure reach their maximum.

For $r>r_{\rm centre}$, the pressure gradient and centrifugal force balance gravity. For $r<r_{\rm centre}$ the centrifugal force increase and balance the gravity and the pressure gradient. Near the ICO, if the inner radius of the torus is smaller than the radius of change of slope of the specific angular momentum profile (Fig.~\ref{fig_Llimitut}), the torus is subject to instabilities. Indeed, the difference between the angular momentum of the tori and the Keplerian angular momentum change faster with radii. This introduces a strong variation of the pressure gradient that balances the gravitation and centrifugal force.

Since the central radius of the torus experiences Keplerian rotation, it should verify $r_{\rm centre}>r_{\rm ICO}$.
The inner radius of the torus is chosen such that the angular momentum of the torus remains always smaller than the maximum allowed angular momentum  $l_{\rm torus}<l_{\rm max}$.

In the case of a rotating boson star, the inner disk radius can be smaller than the ICO. Indeed, the torus must rotate at Keplerian values only at $r=r_{\rm centre}$, so the innermost regions being not Keplerian can survive below the ICO.

The accretion torus is set by choosing the polytropic constant $\kappa$, the polytropic index $\Gamma$,  the torus constant angular momentum $l_{\rm torus}$, the position of the inner torus radius $r_{\rm in}$ and the boson star's  rotation quantum number $k$ and period number $\omega$. With those parameters we can adjust the torus mass, which we define as
\begin{equation}\label{Masstorus}
M_{\rm torus} = 2\,\pi\,\int_{r_{\rm in}}^{\infty}\int_{0}^{\pi}\,\rho\,\sqrt{\gamma}\,\mathrm{d}r\,\mathrm{d}\theta\,,
\end{equation}
where $\gamma=A^4\,B^2\,r^4\,\sin^2\theta$ is the determinant of the spatial metric.

\subsection{Allowed range of the constant angular momentum}
We can classify boson stars according to the possible angular momentum range of their Polish doughnuts in two types.

First, the less relativistic stars ($\omega$ close to $m/\hbar$), which are characterized by a specific angular momentum having monotonic variation with $r$. The smaller angular momentum allowed value for their torus increases with the location of inner torus radius $r_{\rm in}$. If we want the torus to be well-defined everywhere in space, the denominator of Eq.~(\ref{eq_ut_constL}) should be always strictly positive, thus the constant angular momentum should satisfy $l_{\rm torus}<l_{\rm max}$. 
For small values of $\omega$ the allowed value for the angular momentum is shifted to higher values (Fig.~\ref{fig_Llimitut}). This limit value is a function of the inner radius and increase with it (Fig.~\ref{fig_Llimitut}).

Second, the more relativistic boson star (with smaller $\omega$). Fig.~\ref{fig_Llimitut} shows that their Keplerian specific angular momentum does not increase in a monotonic way. Near the ICO, $l_{\rm K}$ has an extremum. Moreover, some of these stars have an ergosphere. In Table~\ref{Tab_cusp}, we present the torus properties in the vicinity of some boson star.
It appears from this study that when the quantum number decreases, the interval of $l_{\rm K}$ where the cusp exist, moves to smaller values and toward the region with the maximum gravitational field.

Regarding spherical boson stars, the two cases we study here in Fig.~\ref{Fig_LlS} show that the Keplerian angular momentum is always smaller than the limiting value for torus and could exist near the centre.

\subsection{Polish doughnuts with a cusp}
\begin{table}
  \caption{
   Existence of a cusp, interval of the specific angular momentum where the cusp may exist, and existence of an ergoregion  for various boson star models, labelled by $k$ and $\omega$.}

  $$
\begin{array}{llllll}
   \hline
   \noalign{\smallskip}
   k & \omega \ [m/\hbar] & 
   \mbox{cusp} &
   \left[l_{\rm min}^{\rm cusp},l_{\rm max}^{\rm cusp}\right] \ [M]& 
   \mbox{ergoregion}\\
   \hline
   \noalign{\smallskip}
   1 & 0.66 &   \mbox{no}& &\mbox{no} \\
   1 & 0.8 & \mbox{no}& &\mbox{no}\\
   2 & 0.535 & \mbox{\mbox{yes}}  & [2.22,2.24] &[0.05,1.01]\\
   2 & 0.8 & \mbox{\mbox{no}} & & \mbox{no} \\
   3 & 0.505 & \mbox{\mbox{yes}} &[2.18,2.25] & [0.06, 1.06]\\
   3 & 0.8 & \mbox{\mbox{no}}  & &\mbox{no}\\
   4 & 0.56 & \mbox{\mbox{yes}} & [2.33,2.5] & [0.15, 0.45]\\
   4 & 0.8 & \mbox{\mbox{yes}} & [3.356,3.334] & \mbox{no}\\
   \hline
 \end{array}
 $$
 \label{Tab_cusp}
 \end{table}

We have mentioned earlier the importance of the existence of an extremum at small $r$ in the profile of the Keplerian specific angular momentum for the most relativistic (smaller $\omega$) boson stars (see Fig.~\ref{Fig_allL}). With such a non-monotonic profile, a Polish doughnut with constant angular momentum can have more than one Keplerian-rotating point. Its angular momentum should then be within the interval $\left[l_{\rm min}^{\rm cusp},l_{\rm max}^{\rm cusp}\right]$ given for various boson stars in Table~\ref{Tab_cusp}.

\begin{figure*}
\centering
\resizebox{4.5cm}{4.5cm}{\includegraphics[width=0.3\hsize]{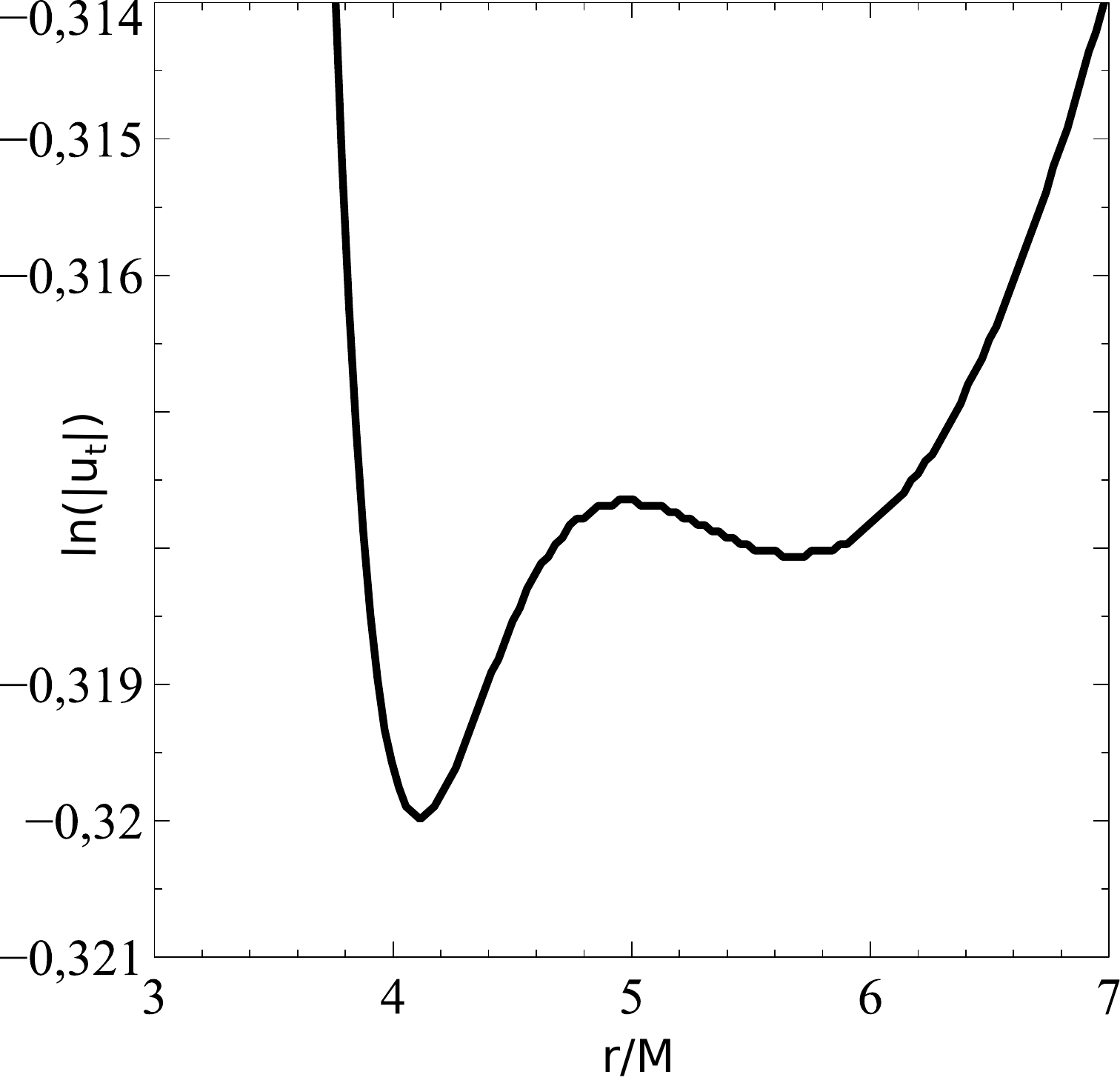}} \hspace{0.5cm}
\resizebox{4.5cm}{4.5cm}{\includegraphics[width=0.4\hsize]{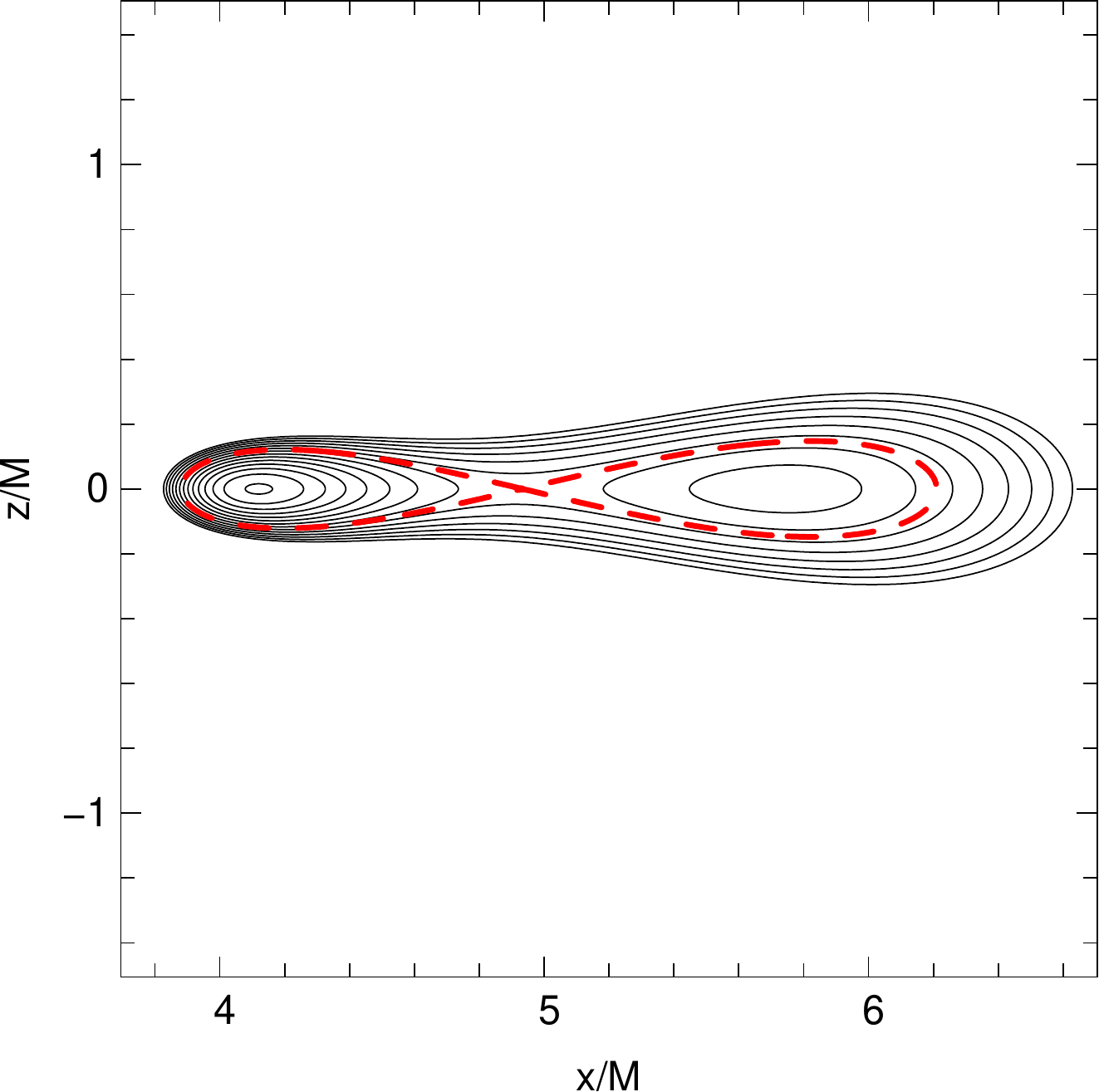}}
\resizebox{4.5cm}{4.5cm}{\includegraphics[width=0.4\hsize]{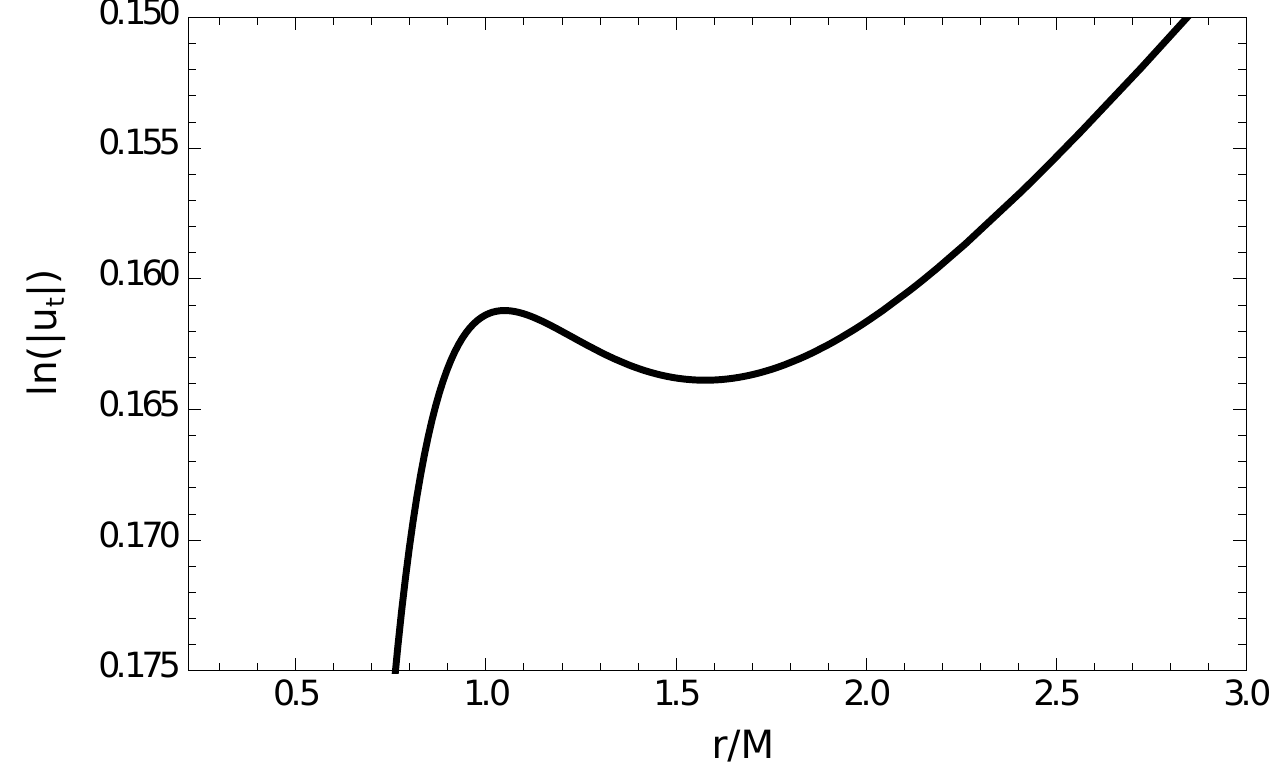}}
\caption{
\textbf{Left:} Profile of the potential $W=\ln(-u_t)$ in the equatorial plane ($\theta=\pi/2$) for a Polish doughnut with $l_{\rm torus}=3.34\,M$ around a boson star with $k=4$ and $\omega=0.8\; m/\hbar$.
\textbf{Center:} Isocontours of $W$ for the same boson star Polish doughnut in the $(x=r \,\sin \theta,z=r\, \cos\theta)$ plane. The self-intersecting potential
line is shown in dashed red, the cusp is the self-intersection point.
\textbf{Right:} Profile of $W$ in the equatorial plane ($\theta=\pi/2$) for a Polish doughnut with $l_{\rm torus}=2.5\,M$ around a Kerr black hole with $a/M=0.9$.
} \label{fig_cusp}
\end{figure*}

It is well-known that Polish doughnuts around black holes exhibit such a feature in particular when their constant angular momentum $l_{\rm torus}$ satisfies $l_{\rm K,ISCO} < l_{\rm torus} < l_{\rm K, mb} $, where $l_{\rm K,ISCO}$ and $l_{\rm K,mb}$ are the Keplerian angular momentum values at the Kerr ISCO and marginally bound orbit. When this condition is fulfilled, the doughnut's potential $W$ has the profile illustrated on the lower panel of Fig.~\ref{fig_cusp}. This profile shows two zeroes of the potential gradient, corresponding to zeroes of pressure gradient, thus to Keplerian rotation. The local minimum of the angular momentum corresponds to the local maximum of the pressure, while the local maximum of $W$ is the so-called torus cusp, a point at which the equipotential surface crosses itself. It is clear that if a torus is thick enough to go beyond the cusp, matter will flow towards the black hole, leading to accretion.

It is interesting to study the equivalent case in a boson star spacetime that also exhibits a non-monotonic variation in the Keplerian angular momentum profile. The upper panels of Fig.~\ref{fig_cusp} show the 1D (in the equatorial plane) and 2D $(x,z)$ profiles of the potential $W$ for a Polish doughnut surrounding a $(k=4,\, \omega=0.8\, m/\hbar)$ boson star with a constant angular momentum $l_{\rm torus}=3.34\,M$. The lower-right panel of Fig.~\ref{fig_Llimitut} shows that this constant value will intersect the Keplerian profile at three radii. These radii  correspond to zero pressure gradient, i.e. zero potential gradient: they appear very clearly in the upper-left panel of Fig.~\ref{fig_cusp}. This potential profile is characteristic for all sufficiently relativistic  boson stars which have a non-monotonic Keplerian profile. Two different values of $r$ may be suitable for being the Polish doughnut's local maxima of pressure: the two local minima of the potential. When the torus is made thick enough to go beyond the ``cusp'' (we keep this name to refer to the local maximum in the potential profile), matter will flow from one centre to the other. It is probable that such a configuration will not be stable (a first numerical simulation is discussed briefly in the next section).

\subsection{Example Polish doughnuts around boson stars}

Our aim here is to present few examples of constant angular momentum Polish doughnuts illustrating cases with and without a cusp. We  also illustrate the location of the accretion torus with respect to the location of the toroidal isocontours of the scalar field. This part is done using the code GR-AMRVAC (Meliani et (2015), in preparation).

We set-up a torus with a central radius  $r_{\rm centre}>r_{\rm  ICO}$, and an inner radius $r_{\rm in}$ that depends on the specific angular momentum of the torus. According to the values of the inner radius $r_{\rm in} $ and the specific angular momentum $l_{\rm torus}$ the torus is either closed or open and reaching infinity. The tori with larger constant angular momentum are open. However, for small values of the angular momentum the torus can be finite in size.

\begin{figure*}
\resizebox{10cm}{8cm}{\includegraphics{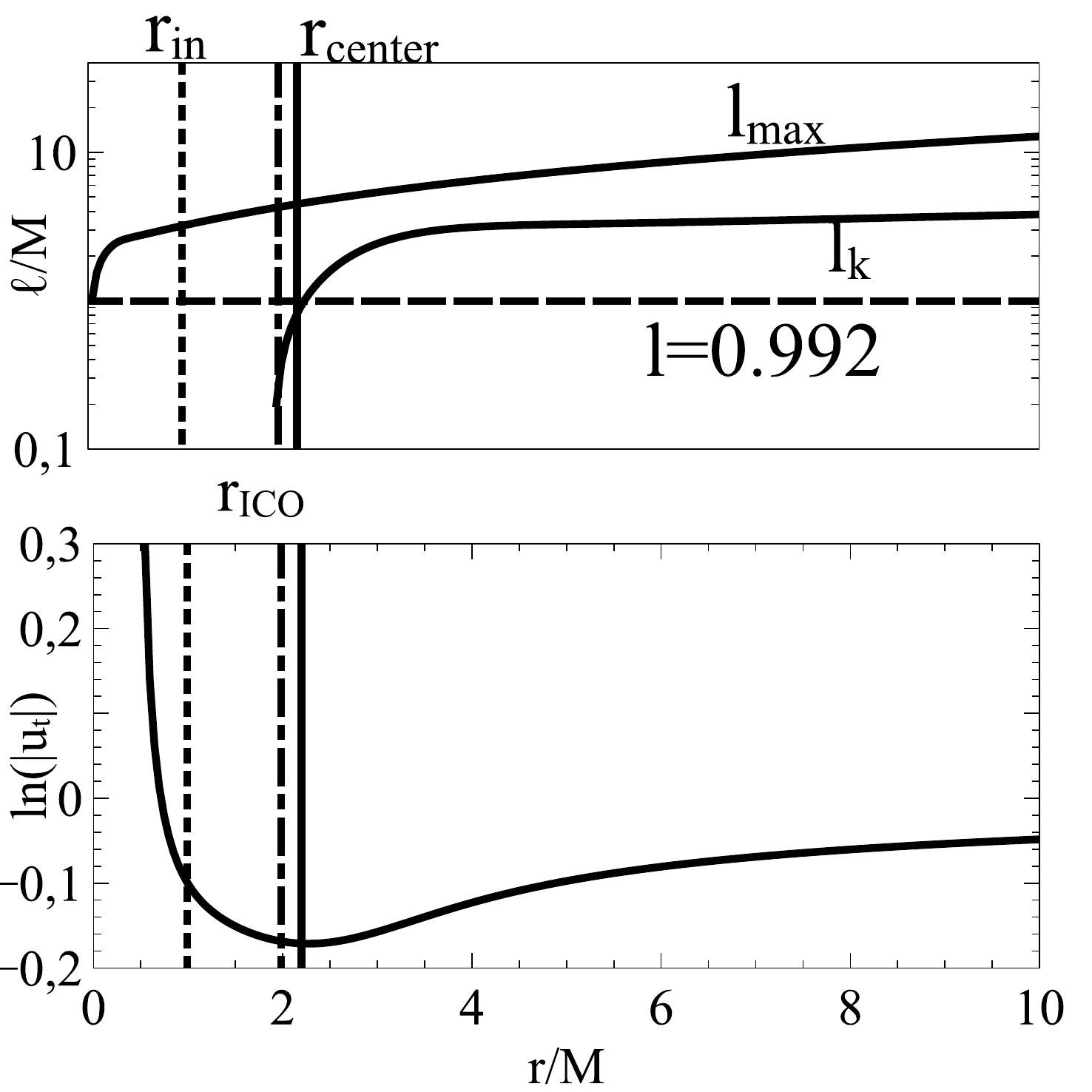}}
\resizebox{6cm}{8cm}{\includegraphics{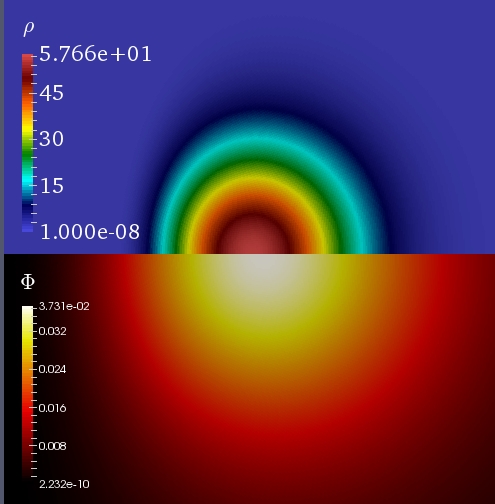}}
\caption{
Torus with $l=0.992\, M$ and inner radius $r_{\rm in} =0.992\, M$ around a boson star with $k=2$ and $\omega=0.8\; m/\hbar$.
\textbf{Left-top:} Keplerian specific angular momentum $l_{\rm K}$ and maximum allowed angular momentum for an $l_{\rm max}$ torus with the marked positions of the ICO, the torus inner radius, and the pressure maximum (labelled $r_{\rm centre}$).
\textbf{Left-bottom:} the torus potential $W=\ln(-u_t)$.
\textbf{Right-top:} the baryon density $\rho$ inside the torus.
\textbf{Right-bottom:} the modulus of the scalar field $\Phi$.
}
\label{Fig_disk1}
\end{figure*}

Let us start with a torus with an inner radius satisfying $r_{\rm in}<r_{\rm ICO}$. The case we show here (Fig.~\ref{Fig_disk1}) has an angular momentum $l_{\rm torus}=0.992\,M$ and an inner radius $r_{\rm in}=0.992\,M$ for a boson star with $k=2$ and $\omega=0.8\; m/\hbar$.  In a strong gravitational field the baryonic torus centre coincides with the centre of the boson star. The structure of the baryonic torus is then congruent with the scalar field profile, since gravitational force points towards the torus centre $r_{\rm centre}$.  The centrifugal force and pressure gradients  support the accretion torus against gravity. At $r<r_{\rm centre}$ the pressure gradient is balanced by the joint gravitational and centrifugal forces, while at $r>r_{\rm centre}$ the pressure gradient and centrifugal force are in equilibrium with gravity. This torus has no cusp and is characterized by strong density and pressure gradients  of the order of the scalar field gradient.

\begin{figure*}
\resizebox{10cm}{8cm}{\includegraphics{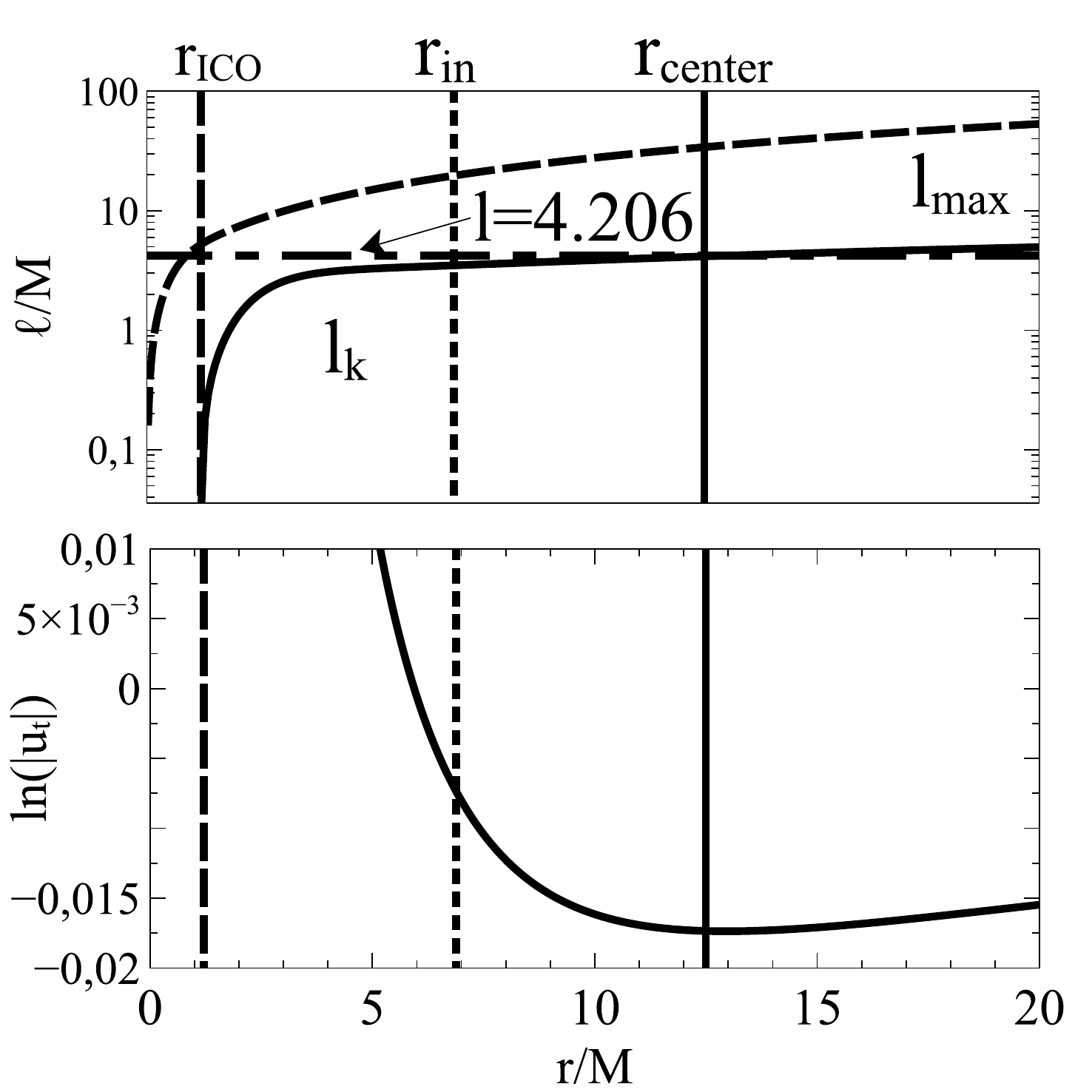}}
\resizebox{6cm}{8cm}{\includegraphics{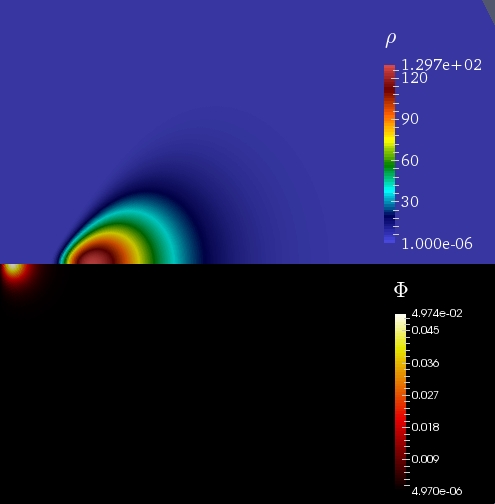}}
\caption{
Same as Fig~\ref{Fig_disk1}, but for a torus with $l=4.206 M$ and $r_{\rm in} =6.88 M$ around a boson star with $k=1$ and $\omega=0.8\; m/\hbar$.
}
\label{Fig_disk2}
\end{figure*}

The second torus has an inner radius satisfying $r_{\rm in}>r_{\rm ICO}$. The case we show here (Fig.~\ref{Fig_disk2}) has an angular momentum $l_{\rm torus}=4.206\,M$ and an inner radius $r_{\rm in}=6.882\,M$, for a boson star with $k=1$ and $\omega=0.8\; m/\hbar$. The torus lies in a region where the scalar field is low and the torus has no cusp.

\begin{figure*}
\resizebox{10cm}{8cm}{\includegraphics{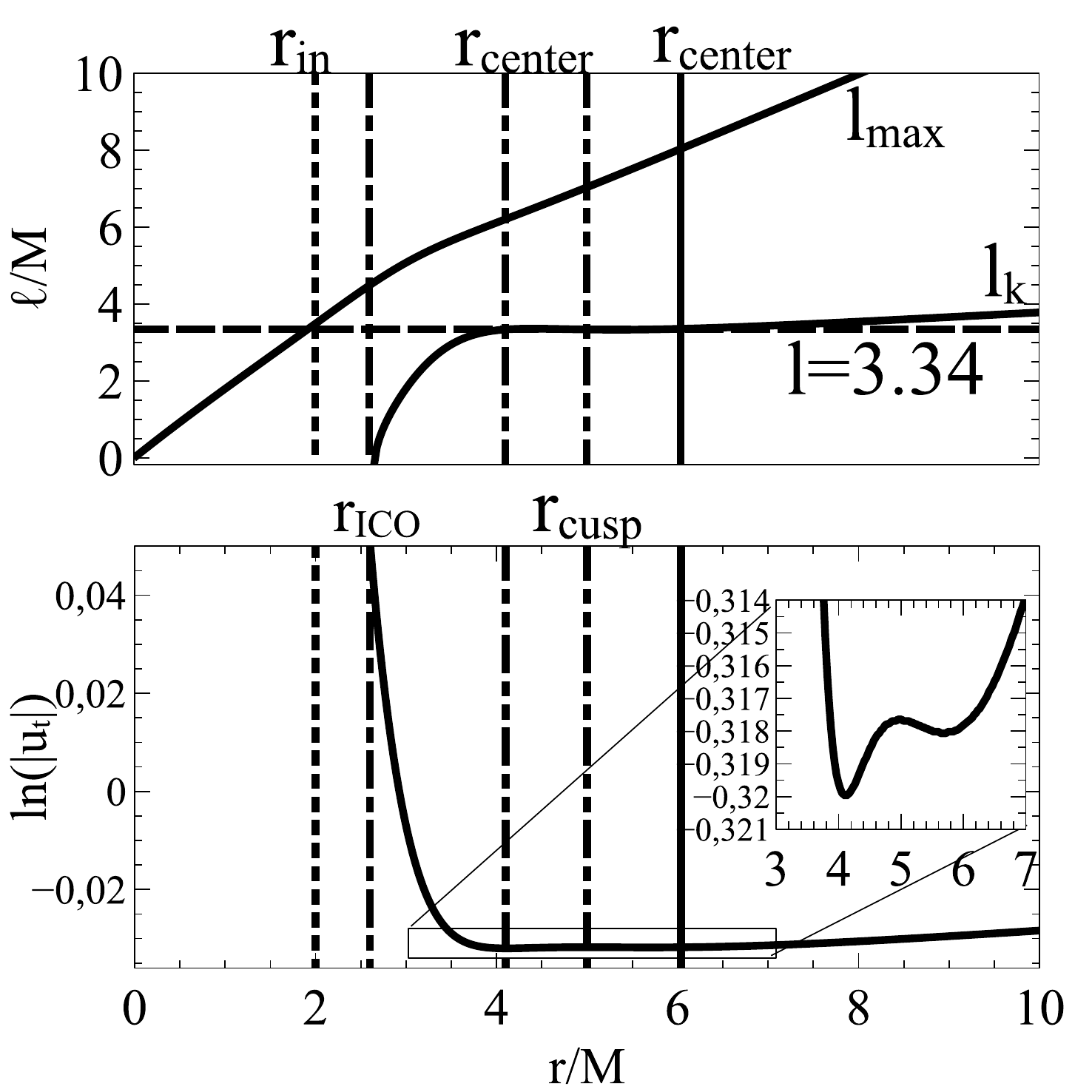}}
\resizebox{6cm}{8cm}{\includegraphics{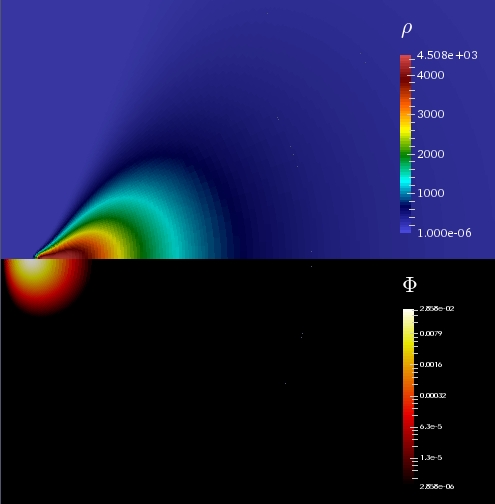}}
\caption{
Same as Fig~\ref{Fig_disk1} but for a torus with $l=3.34 M$ and $r_{\rm in} =2 M$ around a boson star with $k=4$ and $\omega=0.8\; m/\hbar$. Note that there are two local pressure maxima (labelled $r_{\rm centre}$) in this case.
}
\label{Fig_disk3}
\end{figure*}

The third torus (Fig.~\ref{Fig_disk3}) is around a boson star with $k=4$ and $\omega=0.8\; m/\hbar$, has an inner radius $r_{\rm in} = 2 \,M< r_{\rm ICO}$, and a specific angular momentum $l_{\rm torus}=3.34\,M$. This torus has two local pressure maxima and one cusp in between (it has the potential illustrated in Fig.~\ref{fig_cusp}).
With this configuration the matter is attracted to the region where the scalar field reaches its maximum, which is inside the accretion torus.

\section{Conclusions}
We have analyzed the physics of circular orbits around a boson star and investigated a simple analytical model for thick accretion tori characterized by a constant angular momentum (the Polish doughnut model). We found that, except for non-rotating boson stars, circular orbits exist only for $r$ larger than a critical value, $r_{\rm ICO}$, which defines the innermost circular orbit (ICO). Moreover, we found that all circular orbits with $r>r_{\rm ICO}$ are bound and stable. This differs from Kerr black holes where in addition to the stable circular orbits for radii $r \geq r_{\rm ISCO}$ all the way down to the innermost stable circular orbit (ISCO) there exists an unstable circular orbit at the marginally bound radius. Another important difference with respect to black holes is that a boson star is either static ($k=0$) or a fast rotator ($k\geq 1$). In the latter case, many boson stars have $a/M>1$ (Table~\ref{Tab_KOMEGA}), contrary to Kerr black holes.

The study of constant angular momentum tori shows that the specific angular momentum of the torus has a higher limit. The limit value depends on the boson star rotation and pulsation parameters $k$ and $\omega$, respectively, as well as on the inner torus radius. We showed that such tori can exist even if their inner radius lies in the region with maximum gravitational field. This could make a difference in the observed silhouette of the torus surrounding the boson star, with respect to the black hole case. The effect of gravitation on radiation emitted from this region will be stronger.

The analysis of the thick torus model with constant angular momentum showed that these tori have a cusp only when orbiting the most relativistic boson stars. For less-relativistic boson stars, the constant specific angular momentum of the torus is equal to the Keplerian value only in one location (the torus centre) so that there is no cusp present. For highly relativistic boson stars with strong scalar fields, such as $k=4$ and $\omega\le 0.8\, m/\hbar$, an accretion torus can have two local pressure maxima and a cusp for a range of values of the specific angular momentum. Moreover, if a cusp exists, it lies in the region where the scalar field is close to its maximum value.

\section{Acknowledgements}
      Part of this work was supported by the French PNHE (Programme National Hautes \'Energies). ZM  acknowledges financial support from the UnivEarthS Labex program at Sorbonne Paris Cité (ANR-10-LABX- 0023 and ANR-11-IDEX-0005-02)
      EG acknowledges support from the grant ANR-12-BS01-012-01 \emph{Analyse Asymptotique en Relativit\'e G\'en\'erale}.
      FHV acknowledges support from the Polish NCN grant 2013/09/B/ST9/00060.


\bibliography{biblio}

\end{document}